\documentclass[10pt,preprint]{aastex}

\lefthead{Hillenbrand et al.}
\righthead{Y}

\begin{document}

\shortauthors{Hillenbrand et al.}
\shorttitle{1.035 $\mu$m Y-Band}

\title{The Y-Band at 1.035 $\mu$m: Photometric Calibration \\
and the Dwarf Stellar / Sub-stellar Color Sequence}
\author{Lynne A. Hillenbrand, Jonathan B. Foster}
\affil{California Institute of Technology}
\authoraddr{Dept. of Astronomy; MS 105-24; Pasadena, CA 91125}
\authoremail{lah@astro.caltech.edu}
\author{S.E. Persson}
\affil{Observatories of the Carnegie Institution of Washington}
\authoraddr{813 Santa Barbara Street; Pasadena, CA 91101}
\author{K. Matthews}
\affil{California Institute of Technology}
\authoraddr{Palomar Observatory; MS 320-47; Pasadena, CA 91125}
\authoremail{}

\received{}
\revised{}
\accepted{}
\journalid{}{}
\articleid{}{}

\begin{abstract}

We define and characterize a photometric bandpass (called ``Y") 
that is centered at 1.035 $\mu$m, in between 
the traditionally classified ``optical'' and ``infrared'' spectral
regimes.  We present Y magnitudes and Y-H and Y-K   
colors for a sample consisting mostly of photometric and spectral
standards, spanning the spectral type range sdO to T5V.
Deep molecular absorption features in the near-infrared spectra 
of extremely cool objects are such that the Y-H and Y-K colors grow 
rapidly with advancing spectral type especially from late M through mid L,
substantially more rapidly than J-H or H-K which span a smaller total 
dynamic range.  Consistent with other near-infrared colors, however, 
Y-H and Y-K colors turn blueward in the L6-L8 temperature range with later
T-type objects having colors similar to those of warmer M and L stars. 
Use of the Y-band filter is nonetheless promising for easy identification 
of low-mass stars and brown dwarfs, especially at young ages.  The slope of the 
interstellar reddening vector within this filter is A$_Y$ = 0.38 $\times A_V$. 
Reddening moves stars nearly along the YHK
dwarf color sequence making it more difficult to distinguish 
unambiguously very low mass candidate brown dwarf objects from higher mass 
stars seen, e.g. through the galactic plane or towards star-forming regions. 
Other diagrams involving the Y-band may be somewhat more discriminating.

\end{abstract}

\keywords{instrumentation: photometers -- stars: general --
stars: low-mass, brown dwarfs -- ISM: extinction -- infrared: stars}

\section{Introduction}

The traditional near-infrared photometric bandpasses that are observable
from the ground through windows of moderate atmospheric transparency 
are JHKLMNQ (and their slight variants K', Ks, L', M', N1, N2, N3, etc).  This
set of filters defined broadly as in Johnson (1966), with the H-band 
being devised soon thereafter by Becklin and similarly the Q-band by Low,
neglected to include one rather clean window through the atmosphere 
near 1.035 $\mu$m.  This region of the electromagnetic spectrum is 
in between the traditional ``optical'' and ``infrared'' wavelength regimes,
which are usually divided from one another based on detector technology 
(nowadays CCDs vs InSb or HgCdTe arrays) around 1 $\mu$m.  
We discuss in this paper a filter called ``Y'' that takes
advantage of this window and has particular interest for studies of
low-mass and extremely cool stars and substellar objects, i.e. brown dwarfs.

The Y filter bandpass was designed by S.E.P. and K.M. after visually inspecting 
the USAF's LOWTRAN (Selby \& McClatchey, 1975) atmospheric transmission 
curve and also referring to Manducca and Bell (1979). The half-power points 
of the filter were aimed at the 75\% transmission points of the broad 
atmospheric absorption features near 0.955 and 1.112 $\mu$m. The filters were 
manufactured by Barr Associates (Worchester, MA). 
We name the filter ``Y'' in publication as a
compromise between its current discrepant designations (X and Z as described
below).  Further, we choose Y to distinguish this filter from z and Z
filters in use which are quantitatively different.

\begin{figure}
\plotone{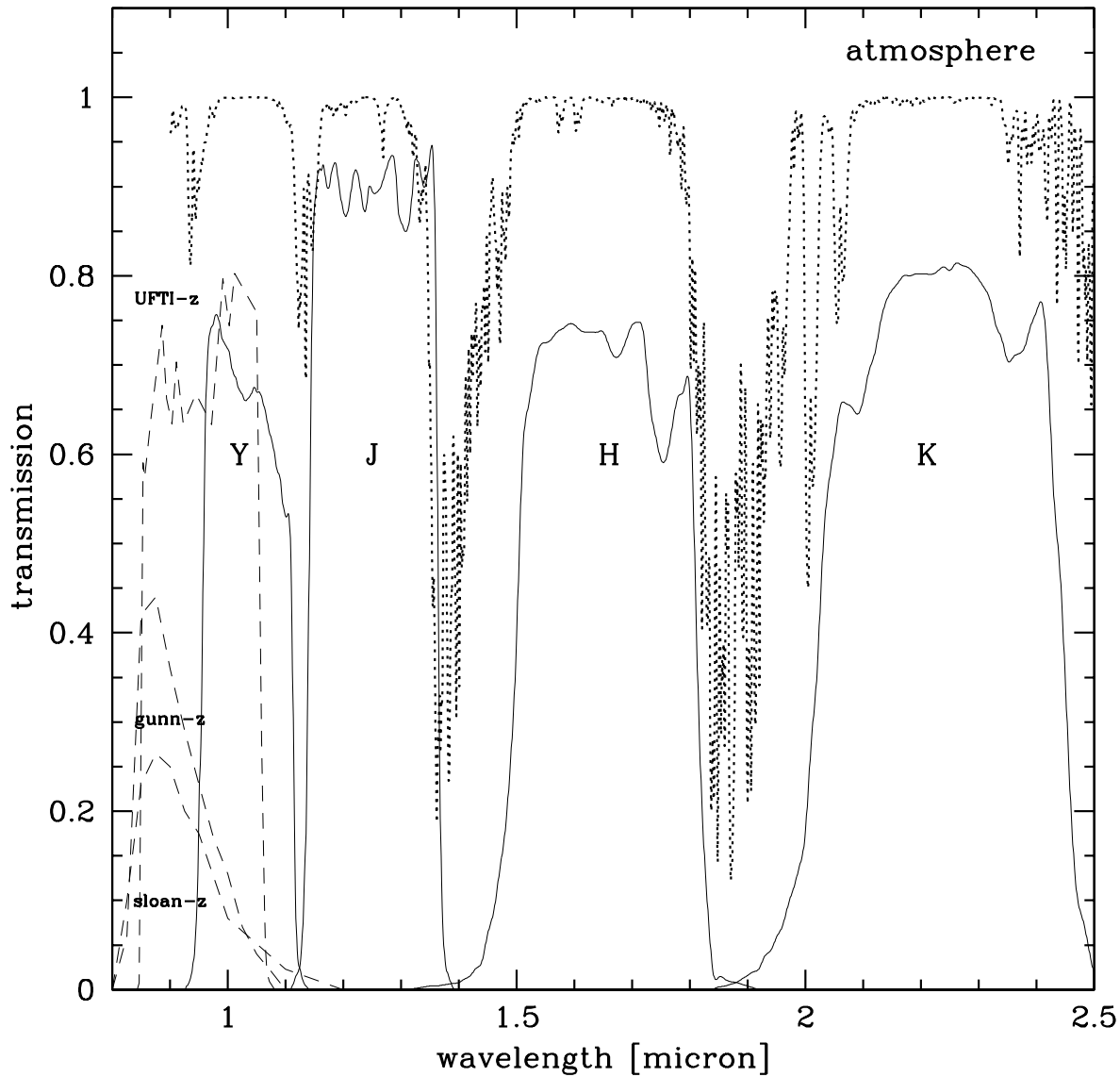}
\caption{Near-infrared Y, J, H, and K filter transmission profiles 
(solid lines) for the Palomar 60" cassegrain infrared camera (P60/IRC)
filters at 77 K, superposed on atmospheric transmission at Mauna Kea 
(dotted line).  Also shown for comparison to the Y-band filter are the
optical Gunn-z and Sloan-z filter profiles convolved with typical CCD 
response and the UFTI-Z filter (cold and already convolved with the Mauna Kea 
atmospheric transmission as presented in Leggett et al, 2002).  
The near-infrared Y is longward of previously defined z and Z filters, and
takes advantage of the relatively clean
atmospheric window centered near 1.035 $\mu$m.
}
\label{fig:atmos}
\end{figure}

Nearly identical Y filters reside in the
Keck Near-InfraRed Camera (NIRC, where it is called ``Z") 
and the Palomar 60" cassegrain InfraRed Camera (P60/IRC, where it is called 
``X"), as well as in the Palomar 200" D-78 and D-80 cameras.  
At 77$^\circ$ K, the Keck/NIRC filter 
has $\lambda_c=1.032\mu$m and $\delta\lambda_c=0.157\mu$m 
with $>$50\% transmission from  0.965 - 1.115 $\mu$m
while the P60/IRC filter has
$\lambda_c=1.04\mu$m and $\delta\lambda_c=0.15\mu$m 
with $>$50\% transmission from 0.9540-1.1105 $\mu$m.
The two filter profiles are nearly identical in shape with the Keck/NIRC
filter 3-5\% less transmissive.  
It was  found to be difficult to produce the specified filter with peak	
transmission much above 70\%;  transmission profiles are given in Table 1. 
In Figure~\ref{fig:atmos}, 
filter transmission curves for the P60/IRC filter set
are shown relative to a model 
of the atmosphere above Mauna Kea, produced using the program 
IRTRANS4 as presented on the UKIRT worldwide web pages 
\footnote{http://www.ukirt.jach.hawaii.edu/JACpublic/UKIRT/astronomy/calib/atmos-index.html.}
for an altitude of 4200 km
assuming 1.2 mm precipitable water vapor and zenith pointing. 
The Y-band is a relatively clean atmospheric window,
both in terms of molecular absorption (Figure~\ref{fig:atmos}) 
and in terms of O$_2$ and OH night sky emission lines 
which are insignificant in the Y-band compared to J-band and H-band.
Thermal background at Y-band is also insignificant.

The only other filter located in this atmospheric window is,
ironically, the F1042M medium/narrowband filter in WFPC-2 aboard HST which has
$\lambda_c=1.015\mu$m and $\delta\lambda_c=0.03\mu$m.
Since the original preparation of this manuscript we have also become aware 
of the UFTI ``Z" filter as described in Leggett et al. (2002).
This filter has half-power points at 0.851 and 1.056 $\mu$m (their Table 2), 
slightly blueward of and broader than our Y filter,
and inclusive of the strong and broad atmospheric water absorption around 
0.93 $\mu$m (see Figure~\ref{fig:atmos}).
The optical ``Gunn z'' and its successor the ``Sloan z'' filters
(Fukugita et al. 1996) as well as the ``Bessel z'', by contrast, are defined 
as long-pass filters and hence have their effective profiles determined by
a rapidly falling CCD response curve towards 1 $\mu$m 
(Figure~\ref{fig:atmos}).
The effective bandpass of these optical z filters 
is centered near 9250 $\AA$ but
can vary substantially from instrument to instrument depending on the CCD
(e.g. SITe detectors vs the new red-sensitive LBNL devices 
as compared in NOAO Newsletter 67, p. 3).

The Y filter introduced here has a well-defined bandpass, as opposed to the 
long-pass optical z filters, and takes advantage of 
the relatively clean atmospheric window centered near 1.035 $\mu$m,
in moderate contrast to the UFTI Z filter.  
Y-band photometry therefore should be relatively independent of the 
specific detector and relatively insensitive
to site-to-site differences in the transparency vs wavelength
and the night-to-night variability of water vapor at a given site.

\begin{figure}
\plotone{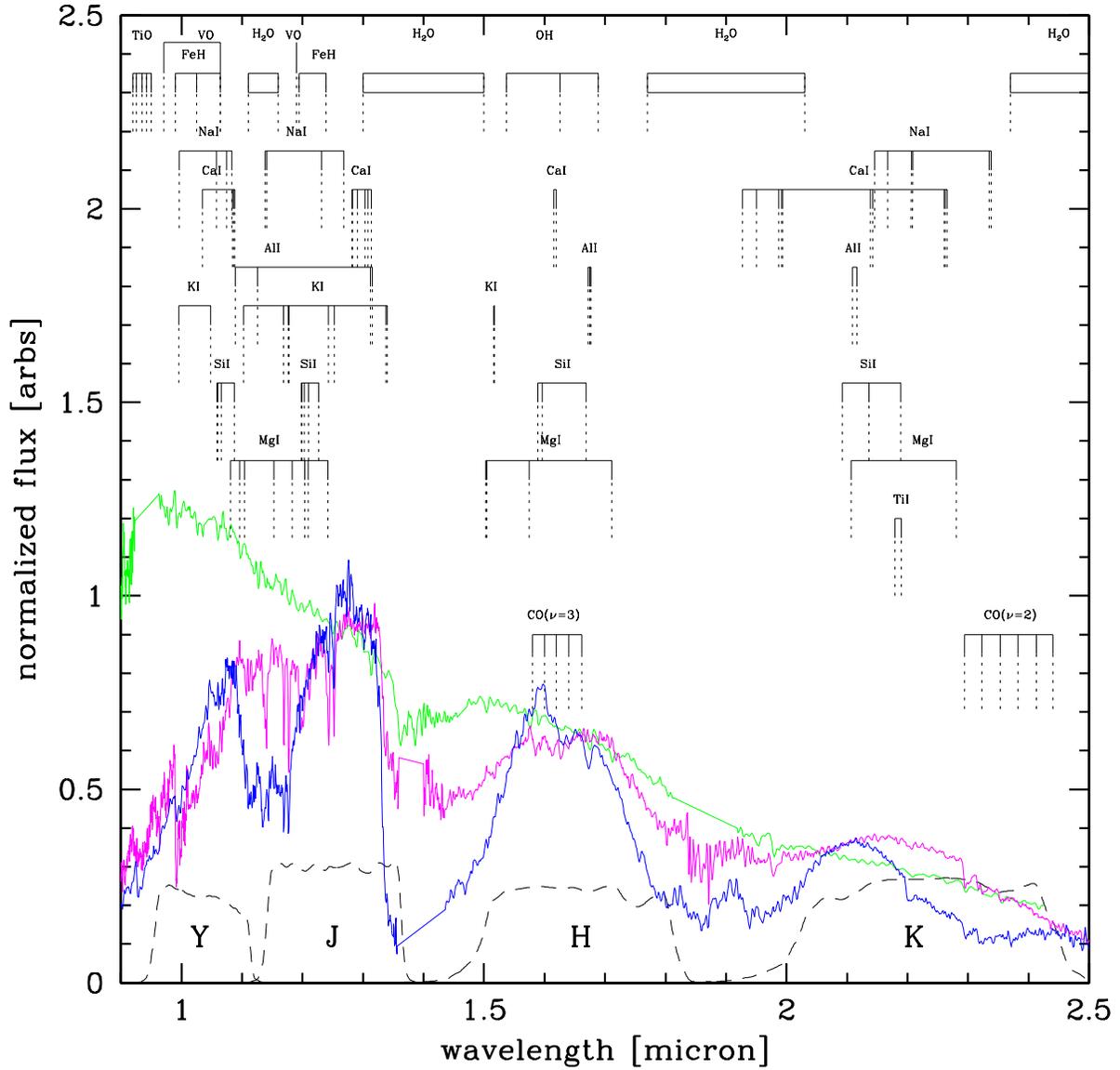}
\caption{Near-infrared Y, J, H, and K filter transmission profiles 
(same as in Figure 1) superposed on the near-infrared spectra of
LHS 386 (M1V; top), 2MASS 0746 (L0.5V; middle), and SDSS 1254 (T2V; bottom)
from Leggett et al. (2000ab) and Geballe et al (2002). 
Top/middle/bottom refers to flux in the H$_2$O bands in between the filters;
the spectra are arbitrarily normalized at 2.1 $\mu$m.
Contributors to spectral absorption features are identified.
The near-infrared Y-band is located near the peak flux of stars $\sim$3000 K
in temperature ($\sim$M4.5-M5V in spectral type).  The peak flux shifts into
and remains in the J-band at coolor temperatures down to 500 K (Burrows et al.,
2001) suggesting that Y-J colors may continue to increase
relative to Vega's zero color even though Y-H and Y-K (in addition to
J-H and H-K) colors turn blueward for T dwarfs.
}
\label{fig:star}
\end{figure}

The motivating forces behind the present study are
the instrumental/atmospheric advantages of the 1.035 $\mu$m window 
just mentioned, combined with the peak brightness of cool and ultra-cool dwarfs
slightly longward of 1 $\mu$m.  As shown in Figure~\ref{fig:star},
wavelengths $\sim$1$\mu$m represent the absolute peak
in the spectral energy distributions of low-mass stars
with spectral types $\sim$M5, 
and a strong local peak for stars and young brown dwarfs with spectral
types $>$L6 (Burrows et al. 2001).  In these very late type dwarfs
the atmospheric opacity redistributes flux far from
expectations based on blackbody radiation (Marley et al., 1996). 
Bolometric corrections become {\it positive} 
in the near-infrared bands, and J-H and H-K colors become 
{\it bluer} with decreasing effective temperature (Burrows et al., 1997).  

Our goal here is to investigate the behavior of near-infrared colors involving 
the Y-band as a function of spectral type, and to determine the slope of 
the interstellar reddening vector.  Our specific interest is in assessing
the utility of the Y-filter in studies of low-mass objects in 
star-forming regions.

\section{Observations}

Imaging observations were obtained by L.A.H. on the nights of 
1999, July 30 and August 1 (UT) 
using the cassegrain InfraRed Camera (P60/IRC; Murphy et al. 1995)
at the Palomar 60" telescope under photometric sky conditions.
The instrument contains a 256$\times$256 HgCdTe NICMOS-3 array 
with 40 $\mu$m pixels and produces a final platescale of 0.62"/pixel.  
Observations at Y (called ``X" in P60/IRC), H, and K 
were obtained using a 9-point dither pattern 
centered on target with 10" separation between adjacent grid points.

Dwarf optical standard stars spanning a range of B-V color 
from -0.32 (spectral class O) to 2.2 (spectral type M6.5)
were selected from Landolt (1992) and supplemented at the red end 
with late-type objects from Kirkpatrick, Henry, \& McCarthy (1991),
Tinney (1993), Kirkpatrick, Henry, \& Simons (1995), Reid, Hawley, \& Gizis
(1995), Hawley, Gizis, \& Reid (1996),
Kirkpatrick et al. (1999, 2000), and Burgasser et al. (2002).
The latest spectral type observed was L8V.
Infrared standard stars used for calibration of the H- and K-bands
were selected from Persson et al. (1998) and include both unreddened G-type 
stars and reddened objects of unknown spectral type.
A set of 8 A0 stars was observed in order to define the instrumental offsets
between the Y filter and the H and K filters.  The star
GSPC S813-D (Lasker et al., 1988)
was used to map the atmospheric extinction curve at Y-band.  
Finally, a moderately reddened field ($A_V$ = 10-20 mag) in the 
Ophiuchus dark cloud was selected from the large area survey of 
Barsony et al. (1997) in order to study the interstellar extinction curve
at Y-band.  On 2001, September 9, some additional data for this program 
was obtained using the same equipment, by A. Burgasser and M. Brown, 
focussing on T dwarfs.

Also, observations that pre-date those above were made of
6 late-M- and L-type dwarfs using the Keck I telescope and NIRC 
(Matthews \& Soifer 1994) by L.A.H. and J. Carpenter on 
1999 February 10 (UT).  Details are given in Hillenbrand \& Carpenter (2000).  
Data of relevence to the present discussion were taken using
a 5-point dither pattern in the Y (called ``Z" in Keck/NIRC) and K filters 
with appropriate flat-field and dark current calibration frames also obtained.

\section{Data Reduction}

Our data reduction steps were performed within IRAF and
include the standard procedures of applying
a linearity correction, masking bad pixels, flat-fielding for correction
of sensitivity variations across the array, sky-subtracting, and, finally,
mosaicking the dither pattern to produce a final image for photometry.
We describe these steps in detail for the P60/IRC observations though the same
basic procedures were followed for the Keck/NIRC data.

Non-linearity of the detector array was assessed from a series of 2 to 20 
second exposures of the back of the telescope's primary mirror cover 
taken at the end of the observing run. 
A central section of the array (1/4 of the total area) 
that is relatively defect-free was used to measure ADU vs exposure time.  
The count rate became significantly non-linear above 36,000  ADU, but to be 
safe images with a maximum pixel value above 30,000 ADU were discarded. 

Below 30,000 counts the response was almost, but not quite, linear. 
The advertised correction (on the Palomar 60" web-page) linearizes data 
according to:
$$xl=x+x^{2}*1.85\times10^{-6}*(1+\delta/EXPTIME)$$
where $xl$ represents linear ADU and $x$ raw ADU.
For our data, $\delta$ = 3.385 for an array speed of 1000 msec and 2.6
for a speed of 743 msec; 
hence integration times shorter than 5 seconds were avoided during the 
observations (sometimes by slightly defocussing the telescope when the
observing plan called for observations of bright ($<$7.5) stars
and integrations longer than 10 seconds preferred. Different linearization 
factors were examined and the correction with the lowest chi-squared 
relative to a linear curve was 
$$xl=x+x^{2}*2.1\times10^{-6}$$ without a dependence on integration time
for EXPTIME $>$5 seconds.
Both the magnitude of this correction and its difference from
the standard correction given above are quite small.

Raw bias and flatfield frames were used to verify that the gain and 
readnoise of the detector were similar to the values in Murphy et al. 
(gain = 8e$^-$; readnoise = 30e$^-$). 
Bias frames were acquired as 1.365 second exposures
with the filter wheel in the closed position. Ten frames were averaged 
after rejecting among the frames the highest and lowest value for each pixel. 
On each night a sequence of twilight exposures through each filter 
was used to construct flat field images.
The minimum and maximum values for each pixel were rejected and the individual
frames averaged after subtracting the bias image.  The sky flat was then
normalized to have an overall average value equal to unity.  
A bad pixel list was created based on a histogram of the normalized
flatfield images by flagging all pixels below 0.65 and above 1.2.
This list had considerable overlap with the advertised bad pixel list
and the two lists were merged into a single bad pixel mask.

The images of astronomical targets were proccessed 
first by constructing sky images as averages of
the 9 data frames in each dither sequence
and subtracting this sky image from each individual data frame,
then by determining the relative
offsets in fractional pixels between each component of the dither pattern for
the point source, and finally by applying the offsets and combining 
the individual frames with bad pixel masking into a mosaic. 
These steps were performed in the NOAO/IRAF environment.

\section{Photometry and Calibration}

We performed aperture photometry with the IRAF/PHOT task. 
An aperture radius of 12 pixels (15" diameter on the sky) 
was chosen uniformly for all objects after looking at FWHM values (2-4) pixels 
and nearest neighbor distances for the entire dataset.  The bright A0 stars 
which tended to be in crowded fields due to proximity to the galactic plane,
set the maximum aperture size.  
A growth curve of magnitude vs aperture size 
peaked in the range 11-14 pixels, beyond which the photometric errors increased
substantially as the magnitudes decreased for a sky annulus defined from
20-25 pixels.  Only for a few (1-2) stars with FWHM just slightly 
less than 4 pixels (2.5") were the fluxes possibly
under-estimated using an aperture only $\sim$3 times their FWHM.  
The empirical aperture correction for these objects was difficult to find in 
many cases, but it was examined for several of the largest objects and found 
to be $\sim$0.007 mag. Due to the difficulty in determining this number and 
its relatively small magnitude this correction was not applied. 
Uncertainty in the derived instrumental magnitudes was 
underestimated by IRAF/PHOT. A better estimate was obtained by examining the 
photometry of sources in the individual unmosaicked frames where the 
statistical scatter was observed to be $\sim$0.01 mag.  We therefore assumed 
a minimum photometric error of 0.01 mag and
characterized our instrumental error as error$_{IRAF/PHOT}$+0.01 mag.

\begin{figure}
\epsscale{.5}
\vskip -0.3truein
\plotone{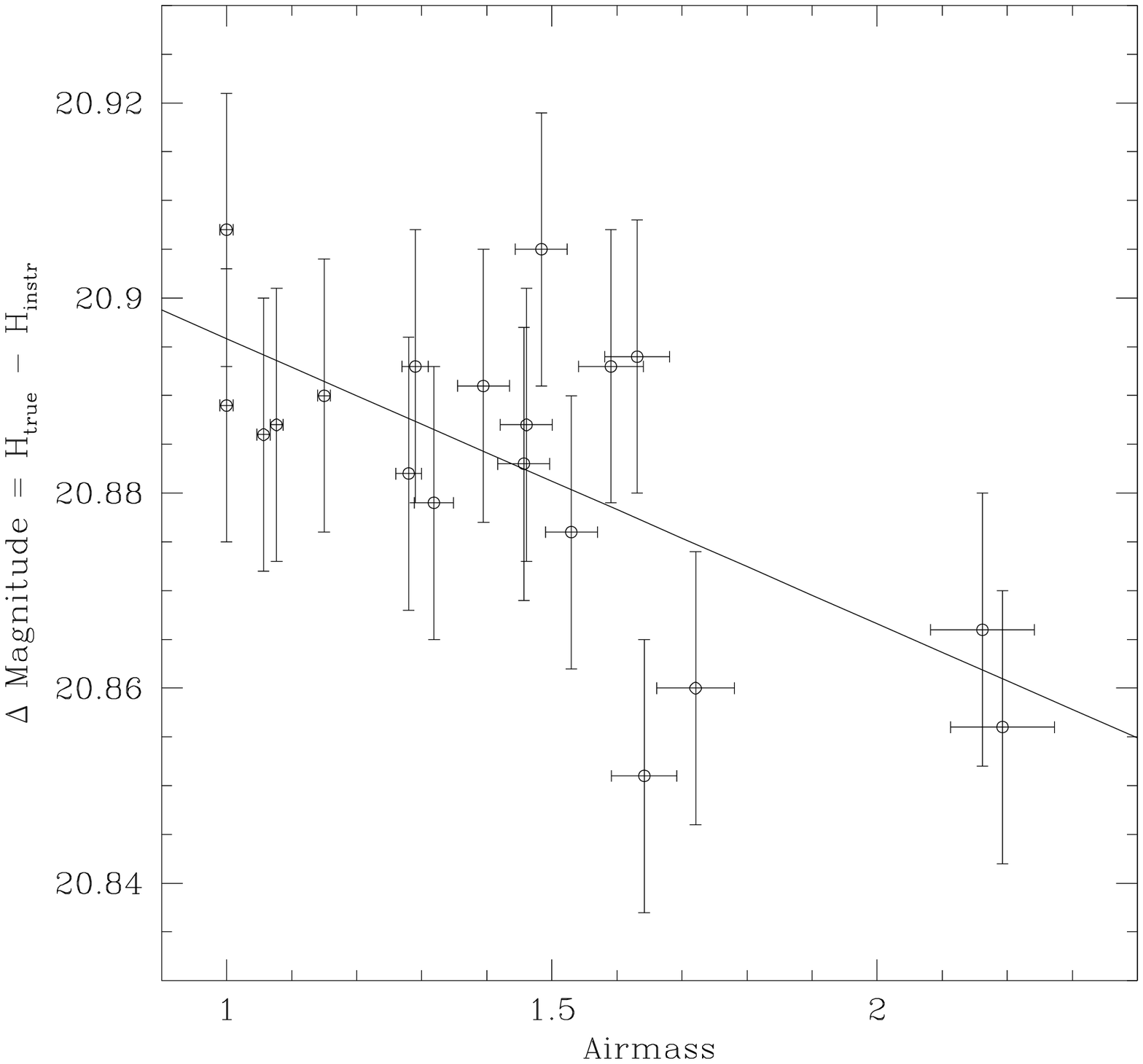}
\vskip -0.3truein
\plotone{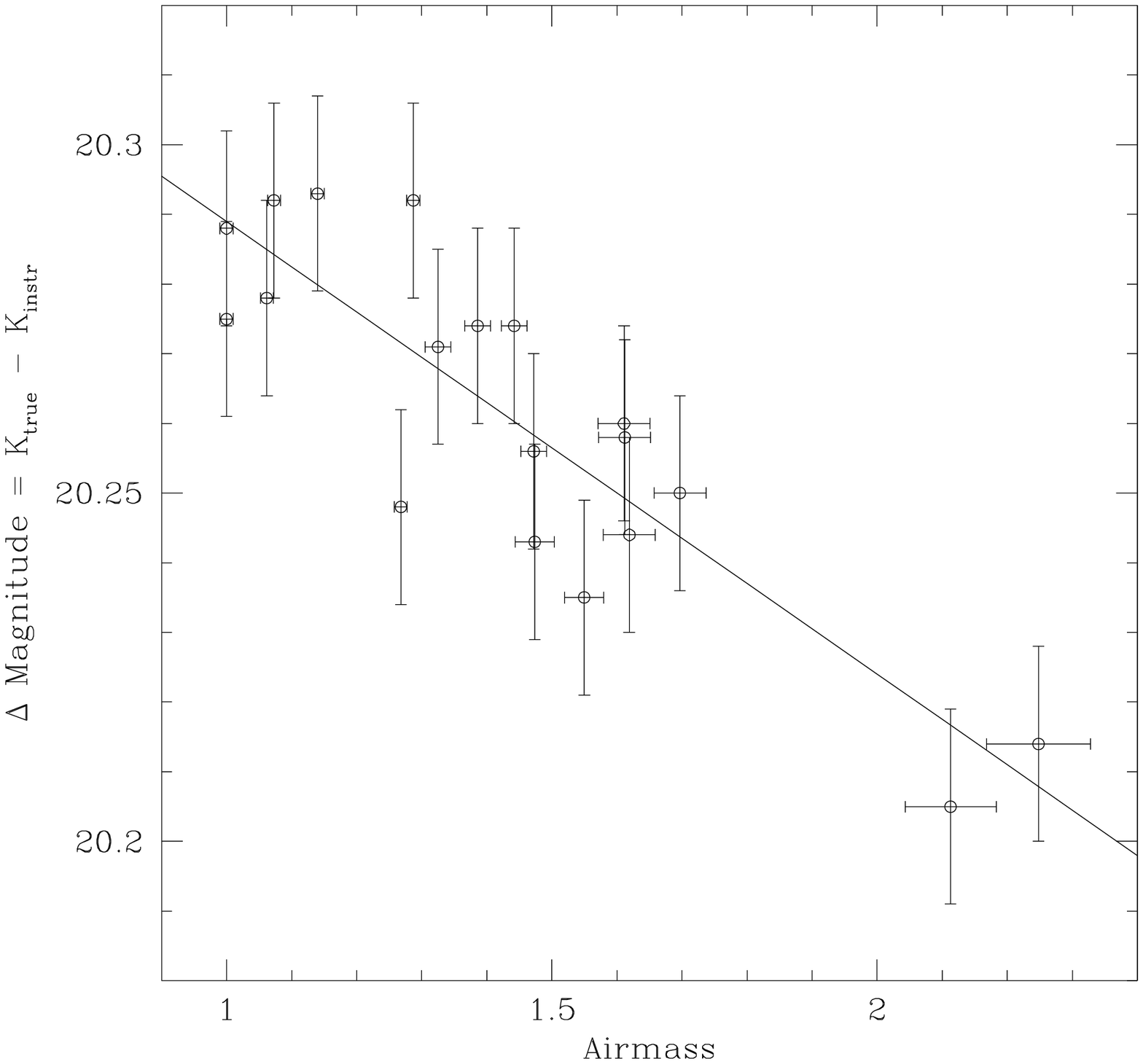}
\epsscale{1.0}
\caption{
Calibration curves for H and K photometry obtained with 
the Palomar 60" telescope and P60/IRC.  See text for 
equations describing the linear fits.
}
\label{fig:airmass}
\end{figure}

\begin{figure}
\plotone{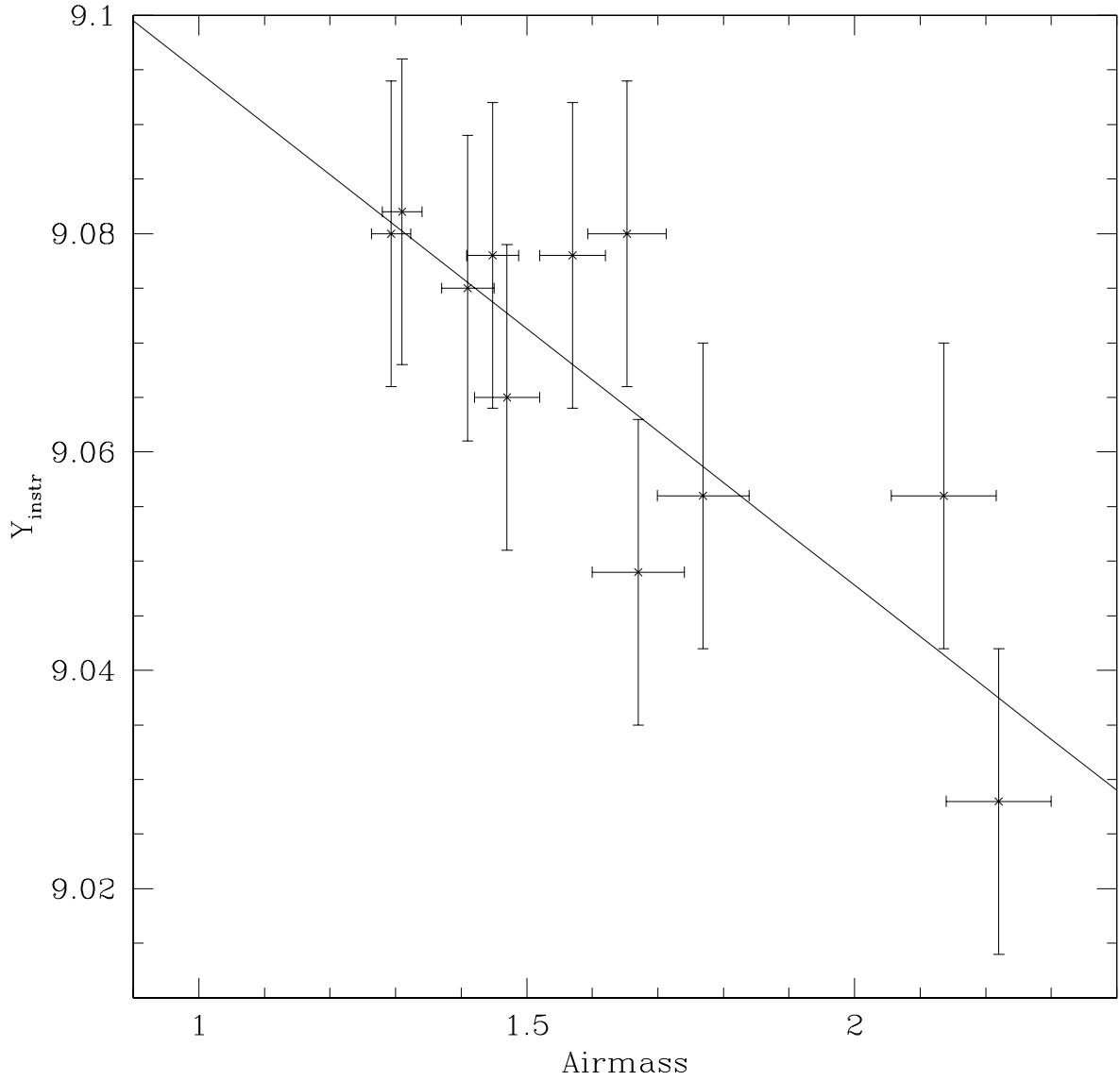}
\caption{
Instrumental magnitude in the Y filter vs airmass for GSPC S813-D. 
Data for both nights on which P60/IRC data were obtained was combined, 
with the second night's values shifted by 0.035 mag to produce
the lowest error in the linear fit.
}
\label{fig:airmass813}
\end{figure}

The relationship between airmass and (standard - instrumental) magnitude 
was used to derive the zeropoints and atmospheric extinction curves 
for the K and H filters (Figure~\ref{fig:airmass}).
Persson et al. (1998) standard stars that were 
observed over the airmass range 1.0-2.5, with airmass taken from the middle 
frame in the 9-position dither pattern, defined these relationships.  
The star GSPC S813-D was observed at 5-6 different airmasses each night 
for the explicit purpose of deriving the atmospheric extinction 
curve for the Y filter without having to know the standard star magnitudes at Y
(Figure~\ref{fig:airmass813}).
Because conditions were photometric,  data from both nights 
were combined in order to obtain a more tightly constrained fit.
We now describe the fitting procedure.

While the slopes from each of the two nights were approximately the same
at Y, H, and K, the zero point for the second night was systematically higher
by a few hundredth's of a magnitude at all wavelengths. 
In combining data from the two nights, various vertical offsets of night
2 data to night 1 data were applied and the value which minimized the $\chi^2$ 
in the linear fit of the extinction curve was retained.  The linear fits 
assumed errors in airmass given by the 1/2 width of the airmass range of
the observations, and errors in the ordinate given by a quadratic sum of
our photometry errors (error$_{IRAF/PHOT}$+0.01 mag) plus an arbitrarily
assumed 0.01 mag error in the standard star magnitudes.  

We thus calibrated the K and H photometry according to:
$$K_{true} = K_{instr.} - (0.065\pm0.010) \times AM_K + 
\left\{\begin{array}{ll}
20.354 \pm0.015(n1)\\
20.389 \pm0.015(n2)
\end{array}
\right\}
$$

$$H_{true} = H_{instr.} - (0.029\pm0.010) \times AM_H + 
\left\{\begin{array}{ll}
20.925 \pm0.015(n1)\\
20.948 \pm0.015(n2)
\end{array}
\right\}
$$
as shown in Figure~\ref{fig:airmass}.
The second term in each equation
contains a numerical value and error for the atmospheric extinction,
$\kappa_{K,H}$, and the third term values and errors for the zero point,
$ZP_{K,H}$. The extinction coefficients we derive for
Palomar, $\kappa_K$ = 0.065 and $\kappa_H$ = 0.029 agree well with those 
from the 2MASS survey
\footnote{http://www.ipac.caltech.edu/2mass/releases/first/doc/sec4\_8.html}
at Mt. Hopkins a few hundred miles away and a few thousand feet higher,  
$\kappa_{Ks}$ = 0.061 and $\kappa_H$ = 0.031.
Note that our quoted zero points are for A.M.=0, i.e. outside the atmosphere,
assuming linear extrapolation of the fit derived from A.M.$>$1; see Figure 3
of Tokunaga, Simons, \& Vacca (2002) for the (in)appropriateness of this
exptrapolation in various near-infrared bandpasses.

To calibrate the Y filter we assume by definition that unreddened A0 stars have 
zero color.  
The A0 stars observed for this program are lightly reddened due to the 
requirement that they be faint enough to observe with P60/IRC 
(implying K$>$7.5 mag)
which is essentially a minimum distance requirement given the absolute 
magnitude of an A0 star, and hence a minimum reddening requirement given
the properties of the local interstellar medium.
We therefore include an interstellar extinction term in our calibration 
equations. We used both the K-band and the H-band data independently to derive 
the zeropoint at Y, from:
$$-K_{instr.}+Y_{instr.} = ZP_K - ZP_Y - (\kappa_K-\kappa_Y)\times AM \\
- (A_K/A_V - A_Y/A_V)\times A_V$$
and a similar equation extrapolating to the Y-band from H-band instead 
of K-band, with the difference between the two approaches being 
less than 0.015 mag and within the $\sim1\sigma$ A0 star photometry errors.
Y$_{instr.}$ is the instrumental magnitude,  ZP$_Y$ is the zero point, 
$\kappa_Y$ is the slope of the atmospheric extinction curve, and $A_Y$ is 
the interstellar extinction. To find $\kappa_Y$ we combined the data for GSPC S813-D 
from both nights, shifting the night 2 data by 0.035 mag to 
obtain the best fit to a straight line, giving $\kappa_Y = 0.047$. 
Figure~\ref{fig:airmass813} shows this slope determination.
For the interstellar extinction term we use literature or Tycho B-V photometry and assume
A$_V$ = 3.1 $\times$ E(B-V), A$_Y$ = 0.38 $\times$ A$_V$ (anticipating our derivation of
this result below in section 5.2), 
and A$_H = 0.155 \times A_V$, and A$_K = 0.090 \times A_V$ (Cohen et al., 1981).
The two A0 stars with less well-determined B-V photometry (see Table 2)
were left out of the final analysis since they introduced larger scatter
and a systematic shift in the results between the H and K derivations.

With $\kappa_Y$ and $A_Y$ in hand, 
the above equation and an analogous one for H are solved for ZP$_Y$; 
errors are calculated by summing in quadrature the errors from each term 
of the equations. From this solution for ZP$_Y$ we then correct the 
10-12 data points for GSPC S813-D and combine these measurements 
with the A0 points to derive the final Y-band calibration equation:
$$Y_{true} = Y_{instr.} - (0.047\pm0.018) \times AM_Y +
\left\{\begin{array}{ll}
20.835 \pm0.030 (n1) \\
20.875 \pm0.030 (n2)
\end{array}
\right\}.
$$
Although the errors in the final K-band transformations
are lower than those in the final H-band transformations,
the zero point at Y derived from K and from H agree to well within the 
estimated errors, differing only by 0.015 mag. 
The $\chi^2$ values for the fits are 0.8 and 0.7 for
the extrapolation from K and H, respectively, a little less than unity
which indicates a good assessment of errors.
The difference in zero point
between the two nights is 0.040 mag whereas we had shifted the data for S813-D
between the two nights by 0.035 to derive the slope of atmospheric extinction,
all consistent, and suggestive that only a single iteration of the procedure
is needed.

The final Y, Y-H, and Y-K photometry for all objects is presented in Table 2
along with errors and ancillary information.
These measurements are in the natural system of the camera, which should be 
that defined by the Persson et al. (1998) standards set up in part 
using this same camera.  Magnitude uncertainties are given by:
$$\sigma_{magnitude} = \sqrt{\sigma_{ZP}^2 + 
		     airmass^2 * \sigma_{\kappa}^2 + 
		     \sigma_{instrumental}^2}$$
which contains 1-$\sigma$ error terms for the zero point, atmospheric extinction
assuming no uncertainty in the recorded airmass, 
and instrumental errors, $\sigma_{instrumental}$ = ($\sigma_{IRAF/PHOT}$+0.01 mag).
Color errors are assumed to be the root-sum-square of the relevant magnitude 
errors which means that we are ignoring the effects of
any correlated errors in our photometry.
Multiple measurements of the same star were averaged, mostly occurring in the
case of GSPC standards. 
Also given in Table 2 are optical photometry and 
spectral types from the literature, and 2MASS photometry from
the Second Incremental Release for comparison with our H and K measurements.

\section{Analysis}

\subsection{Colors}

\begin{figure}
\epsscale{.6}
\plotone{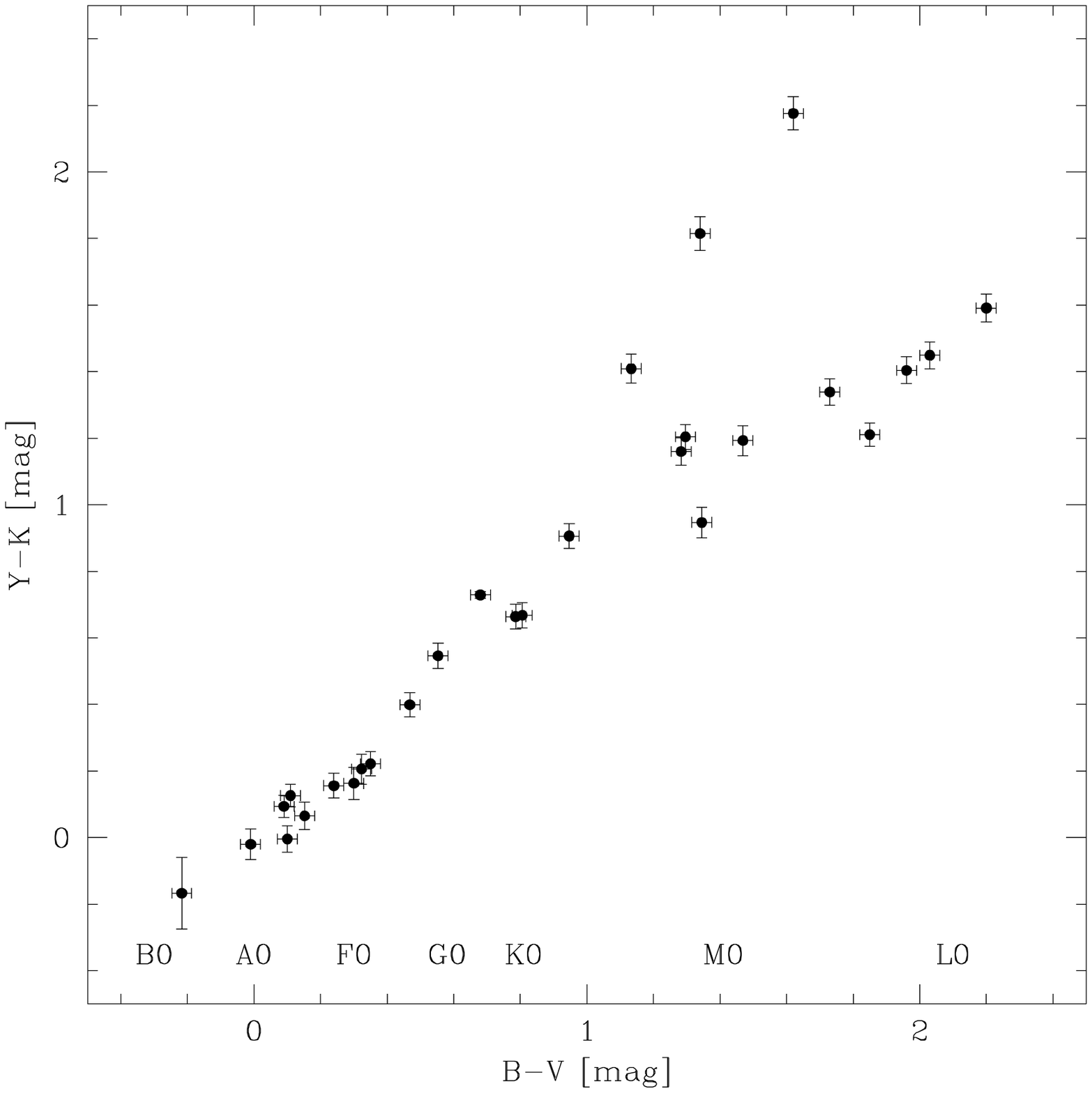}
\plotone{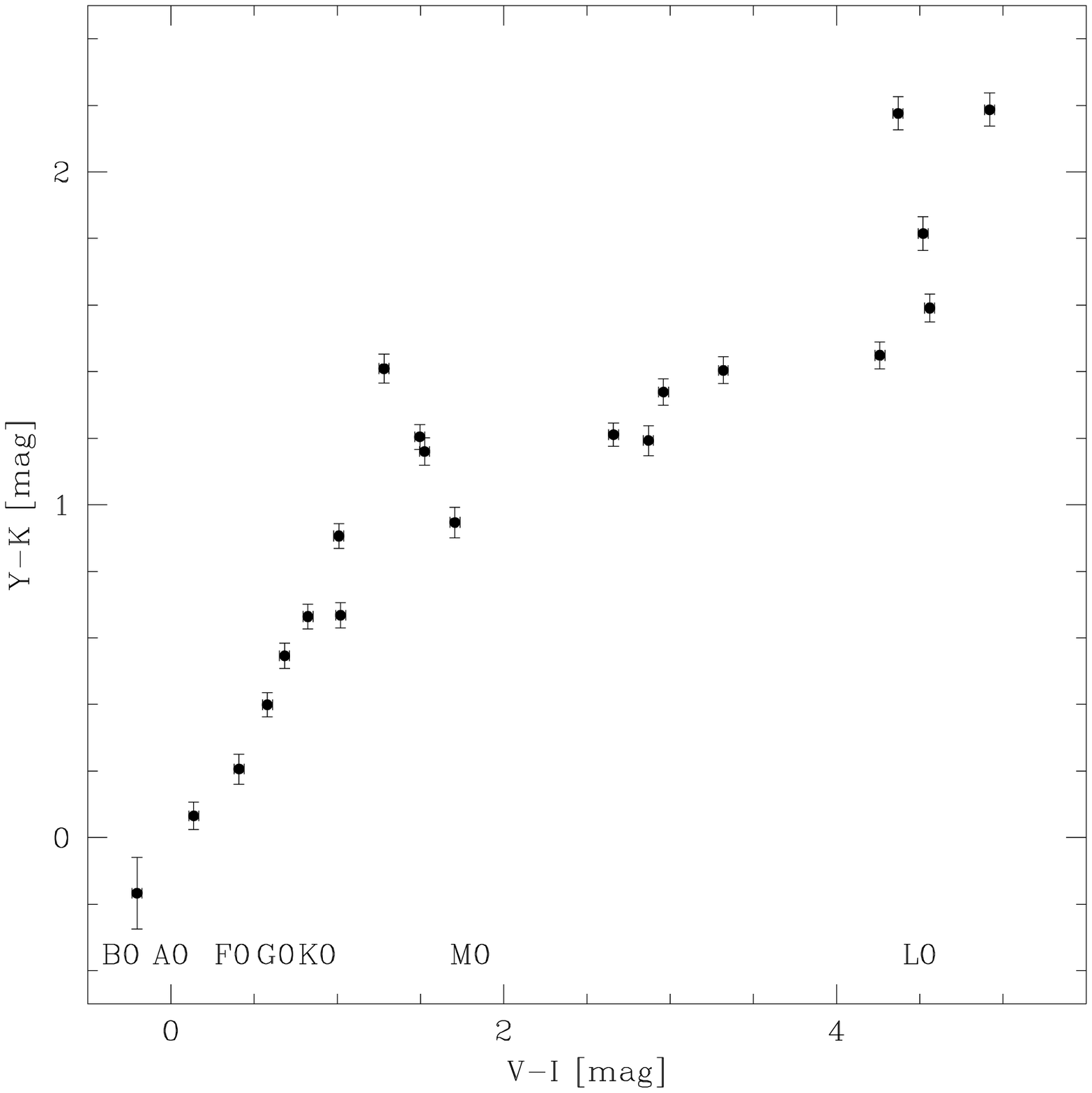}
\epsscale{1.0}
\vskip -0.3truein
\caption{
New near-infrared Y-K colors compared to optical B-V (top panel) and 
V-I (bottom panel) colors from the literature.  
Locations of spectral type labels correspond to optical colors.
The reddest stars in Y-K with published 
optical photometry appearing in these figures are 
2MASSW 1439284+192915 (L1 in the V-I plot only, with Y-K = 2.19) 
LHS 2924 (M9, with Y-K = 2.18), 
LHS 3003 (M7 or perhaps later, with Y-K = 1.82), 
and VB 8 (M7 with Y-K = 1.59).
B-V colors appear to saturate and turn blueward around spectral type M7
while V-I colors saturate around spectral type M8 but do not turn
systematically blueward at late types.
}
\label{fig:optcolors4}
\end{figure}

Color-color diagrams were made for different color combinations using 
the H and K results presented in this paper and B, V, and I photometry
obtained from the literature where available.  These plots were superposed 
on  standard color-color lines (e.g. Bessell \& Brett 1988 transformed to
the CIT photometric system)
and good agreement was seen between our H and K photometry and 
expected I-K and H-K colors.  Moving now to the Y filter,
Figure~\ref{fig:optcolors4} shows the infrared Y-K color compared to
optical B-V and V-I colors.  These colors redden together with increasing
spectral type to mid-M in the case of B-V and mid-L in the case of V-I,
at which point the optical colors become saturated and/or turn bluer while
the infrared color continues to grow redder.
Note that optical colors are not available for all of the objects in Table 2,
particularly the later type, redder stars and brown dwarfs.

\begin{figure}
\plotone{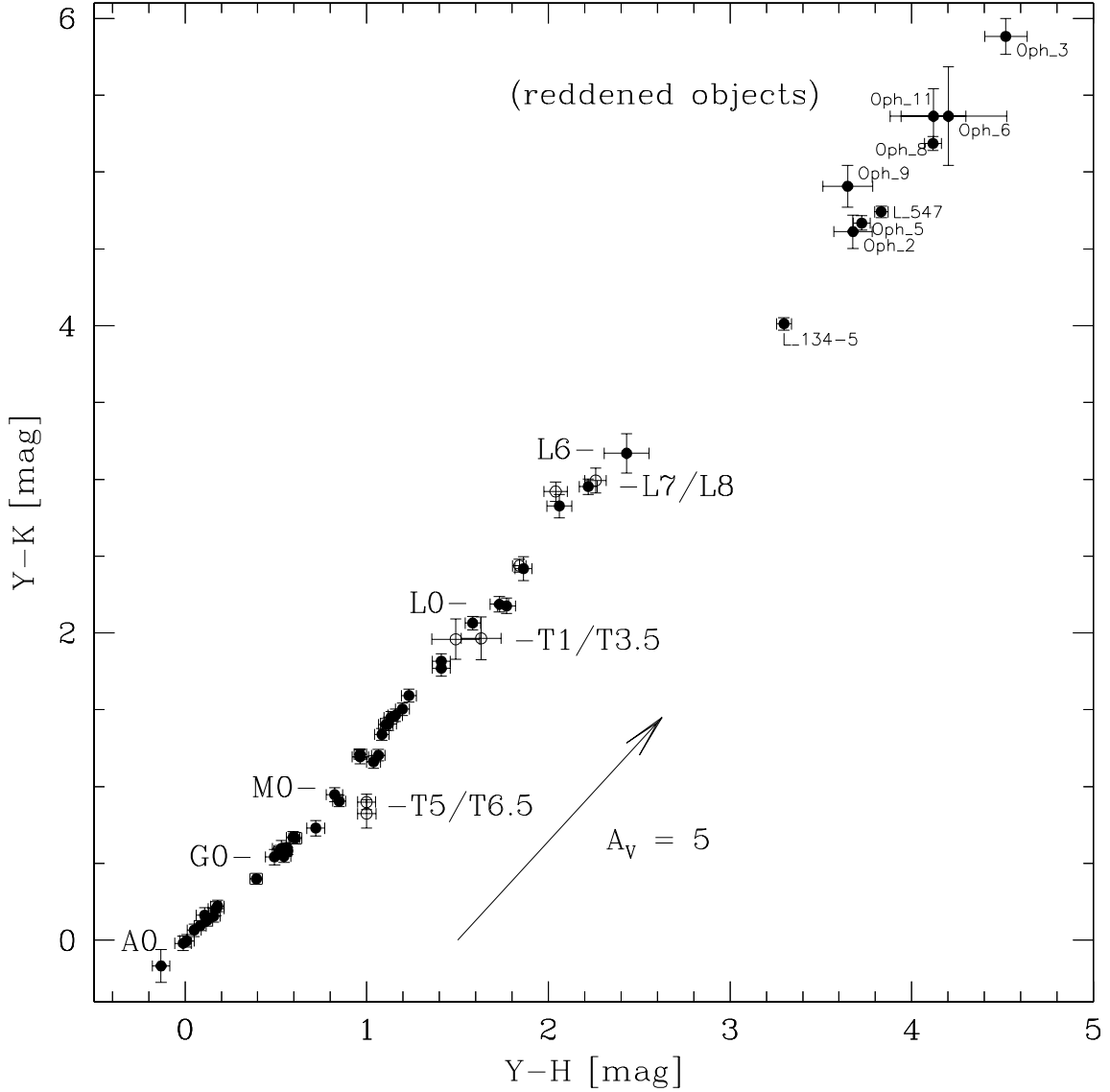}
\caption{
Newly derived Y-K vs Y-H colors for all stars in our sample plus 
Gl229b data from Matthews et al. (1996) taken through similar filters
in the P200/D-78 camera. 
Dwarf colors steadily redden from the earliest type stars through at least 
spectral type L6 stars and brown dwarfs, though become increasingly
bluer from L7 through at least T5.  The spectral type labels correspond to 
Y-K colors; filled symbols
indicate increasing color with advancing spectral type (labelled on the left
of the sequence) and open symbols indicate decreasing color with advancing
spectral type (labelled to the right of the squence).  All objects 
with Y-H and Y-K colors $>3.5$ mag are substantially reddened; the derived
magnitude and slope of the interstellar reddening vector are indicated 
by the arrow.
}
\label{fig:ircolors1}
\end{figure}

\begin{figure}
\epsscale{.6}
\plotone{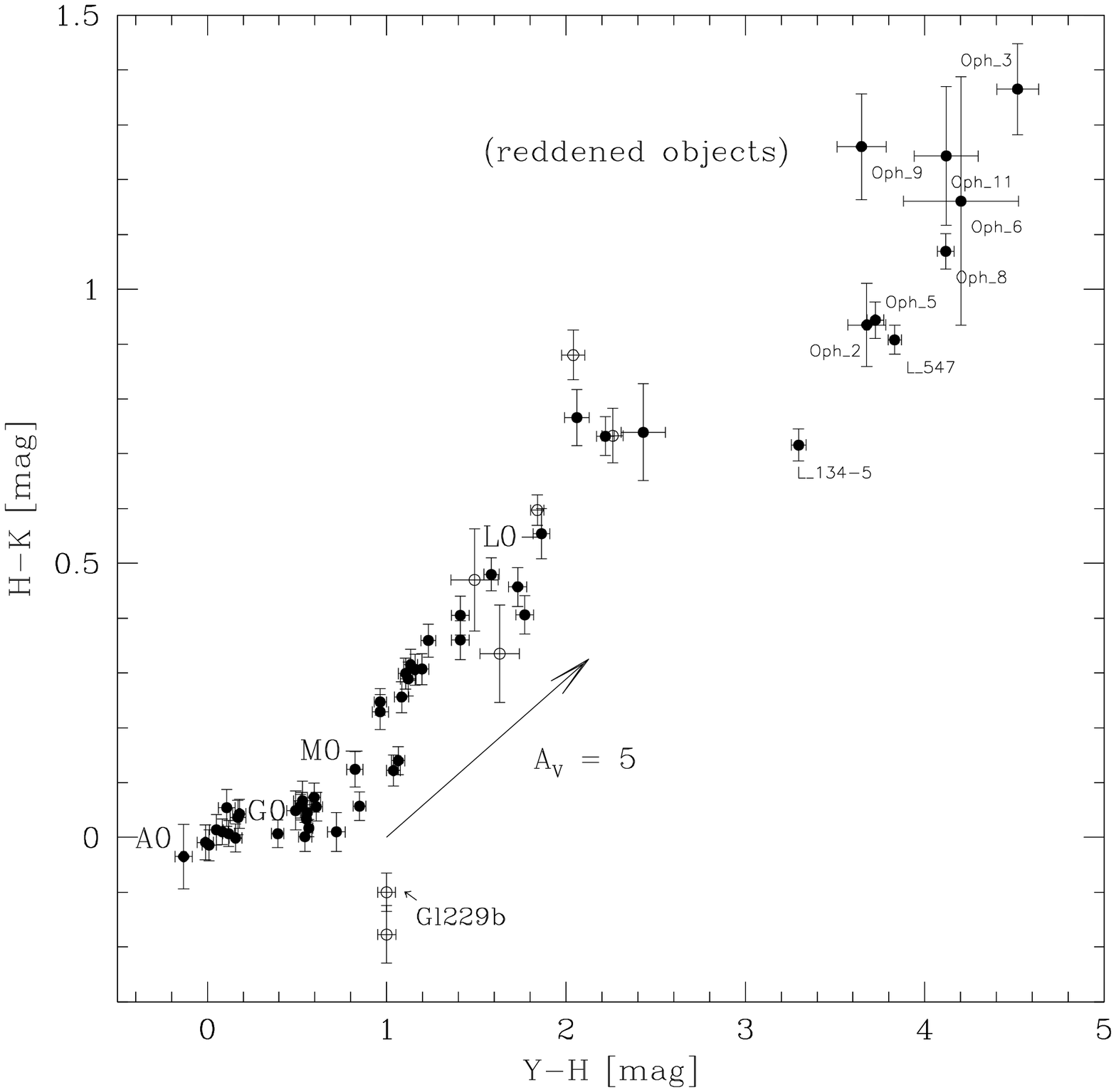}
\plotone{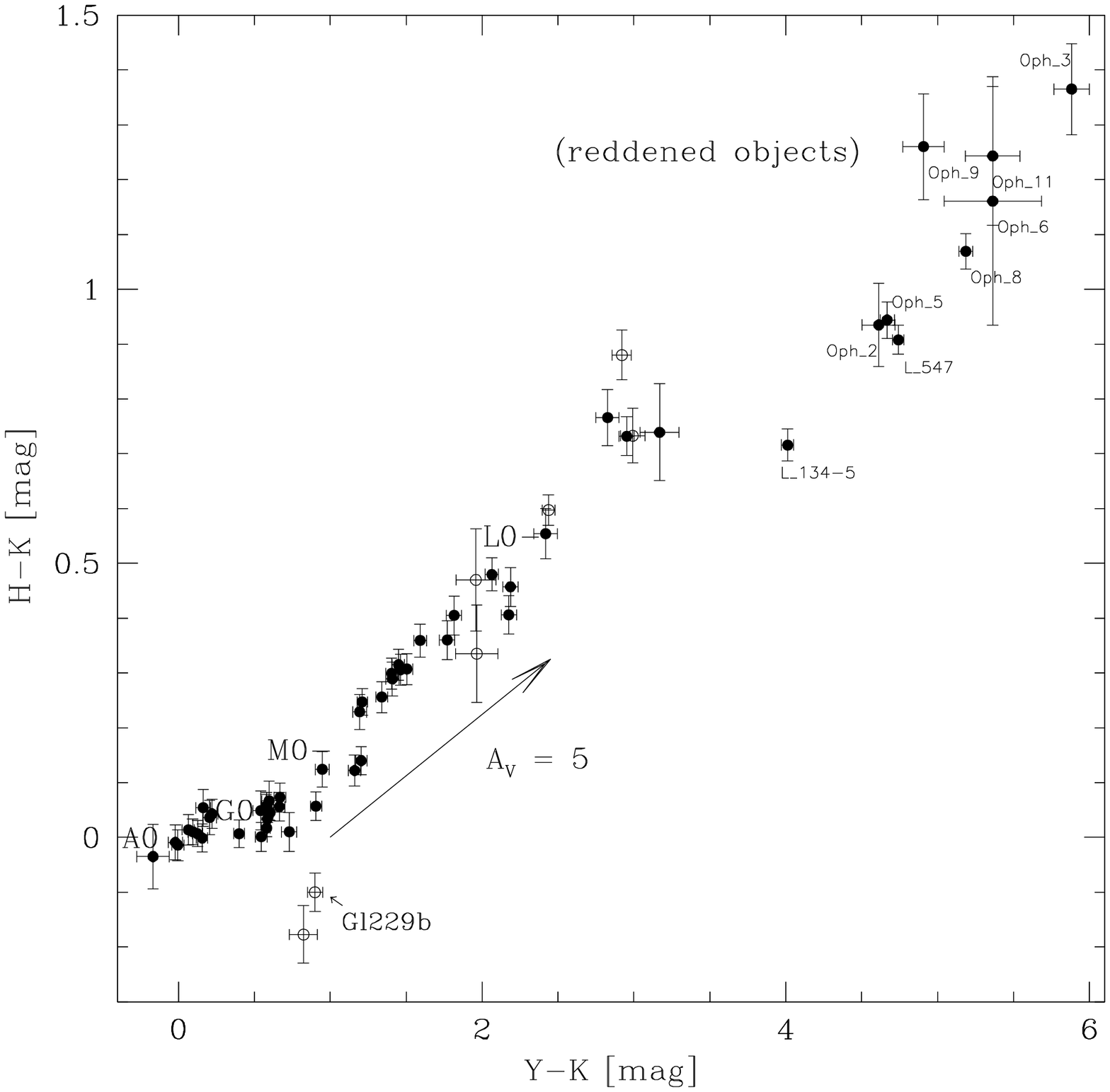}
\epsscale{1.0}
\caption{
H-K colors vs Y-H (top) and Y-K (bottom) colors, similar in other ways
to Figure~\ref{fig:ircolors1}.
The centers of the spectral type labels correspond to H-K colors.
H-K colors are relatively flat through spectral type K7 and
then redden to $\sim$0.75 by L6, flatten to L8, and turn blueward 
at later types.  Y-K and Y-H colors redden more rapidly than H-K, though 
all YHK colors turn blueward by late L and into the T spectral range.
}
\label{fig:ircolors2}
\end{figure}

Plots of Y-H versus H-K, Y-K versus H-K, and Y-H versus Y-K 
were then investigated, as seen in Figures~\ref{fig:ircolors1} 
and ~\ref{fig:ircolors2}.  
The range of spectral types extends from sdO through T5V with photometry for
the T6.5V brown dwarf Gl229b (Matthews et al., 1996) shown for comparison.
Gaps in the data around Y-K = 0.3 and 0.8 mag
are due to the lack of F and K stars in our sample. 
These plots reveal a relatively tight relationship between the YHK colors 
as they grow redder together through spectral type $\sim$L6V.   Beyond L6V
the colors turn blueward and from L7V to at least T6.5V the colors are 
nearly degenerate with those of warmer L and M stars.
%
%
%
%
%
%
%
%
%

There are six objects in common between this study and that of
Leggett et al. (2002): LHS 3003 (M7V), LHS 2924 (M9V), 
2MASS J1439284+192915 (L1V), DENIS-P J0205.4-1159 (L7V),
2MASS 1632291+190441 (L8V), and SDSSp 015141.69+124429.6 (T1) 
Our Y-band magnitudes are 0.3-0.6 mag brighter than their
UFTI Z magnitudes and our colors accordingly bluer for these objects,
consistent with the nature of the filter center and width differences
described in our introduction.  


\subsection{Extinction}

In order to study the effects of interstellar extinction on colors involving
the Y-band filter, a moderately reddened field towards the edge of the 
Ophiuchus dark cloud was observed. Eleven stars were identified in all three 
filter images, but only 7 were bright enough in Y to produce acceptable
photometry.  L~134 and L~547 from Persson et al. (1998) were also included
in this analysis.  The locations of these 9 stars in YHK  
color-color diagrams did not enable us to deduce the magnitude
of the reddening vector, as they lay basically along an extrapolation of
the color-color line defined by the dwarf stellar/sub-stellar sequence
(see Figures~\ref{fig:ircolors1} and ~\ref{fig:ircolors2}). 

In order to quantify how objects redden in Y-band relative to V-band we 
retrieved J, H, and K$_s$ photometry for the 7 Ophiuchus and the 2 Persson 
``red objects'' from the 2MASS survey's Second Incremental Release. 
We then dereddened these 9 stars using the 2MASS J-H and H-K$_s$ colors and the 
Cohen et al.  (1981) interstellar extinction curve until they intersected 
the dwarf color sequence in a J-H vs H-K$_s$ diagram over the color range 
corresponding to spectral types K7-M6. 
Fortunately these stars all appear to be reddened background stars and not
young stellar objects associated with the Ophiuchus (and other, in the case
of the Persson et al. ``red stars'') clouds, 
although we can not prove this.  If they were indeed located 
within clouds they would likely have circumstellar dust/gas and hence colors 
that reflect substantial near-infrared excess in addition to the effects 
of reddening, invalidating our approach.  The estimated H-K and J-H color excess 
due to interstellar reddening was converted to $A_V$.

We then employed our Y-K vs H-K diagram to experiment with different values 
of $A_Y$ and find the value that best put the 9 reddened stars on the 
empirically defined color-color line. 2MASS gives only upper limits to the 
J magnitudes for 2 of the 9 stars, implying lower limit $A_V$'s derived from 
the J-H vs H-K$_s$ diagram; these stars were given little weight in the $A_Y$ 
determination. The relation $A_Y$ = (0.38$\pm$0.03)$\times A_V$ 
was found to best bring the ensemble of stars onto the color sequence defined
by unreddened stars.  The derived value of 
total extinction at Y-band relative to V-band is within the rough expectations
from interpolation between adjacent atmospheric windows, given 
$A_J$=0.265$\times A_V$ and $A_I$ = 0.617$\times A_V$.  
For comparison,
$A_{z(gunn)}$=0.472$\times A_V$  and
$A_{z(sloan)}$=0.453$\times A_V$ have been quoted in the literature
(Schlegel et al. 1998). No information is available to the authors
regarding A$_{Z(UFTI)}$.

\section{Discussion}

Soon after incorporation of the H-band into standard filter sets
it was shown by Bahng (1969) that the colors of GKM stars differ from
blackbody expectations in the near-infrared.  The excursions from monotonic
reddening with advancing spectral type are largely due to the influence 
of H- opacity in the H-band.  Deviation from blackbody colors is apparent
in the now-familiar
J-H vs H-K diagram in the growth of J-H relative to H-K in a manner
that is ``faster'' than blackbody from spectral types G through K, and
then in the peak and blueward turnover of J-H near spectral type K7 as H-K
continues slowly to get redder.  After spectral type M4, J-H again grows redder 
with increasing H-K through $\sim$L6.  

In our data we see that colors involving Y also steadily redden out to 
Y-J $\approx$ 1 mag, Y-H $\approx$ 2.5 mag, and
Y-K $\approx$ 3 mag at spectral type L6V.  Then, Y-H and Y-K colors 
both turn sharply blueward in the L6-L8V temperature range with later T-type 
objects having colors similar to those of warmer L and M stars.  Y-J colors
turn only softly blueward, remaining relatively constant from L6-T4V and 
then overlapping dwarf stars only as warm as M8-M9 by spectral type T5V.

Cool M- and L-type objects initially become redder 
with advancing spectral type because of increased opacity sources at the 
shorter Y-band, mostly H$_2$0 absorption at first, then H$_2$-H$_2$ collision
induced absorption. When CH$_4$ opacity begins to contribute substantially
in the T dwarfs, the H and K fluxes decrease, sending the colors bluer.  
This bluing behavior is consistent with that exhibited in
J-H vs H-K diagrams of late-type objects (e.g. Burgasser et al. 2002)
where the transition between L-dwarfs and T-dwarfs takes colors from the reddest
occupied by stars, all the way though the heart of the color-color diagram 
past every known astronomical object, to the bluest colors occupied by both 
hot stars and cool brown dwarfs.  In the case of YHK colors, however, instead
of crossing the entire color sequence the T dwarf sequence overlaps
the only slightly warmer L and M dwarfs 
(see Figures~\ref{fig:ircolors1} and ~\ref{fig:ircolors2}).  The latest
T dwarfs ($>$T5) appear to fall more blueward in Y-K than Y-H and hence
still may be uniquely distinguishable in this color-color plane.  Overall,
Y-J appears to be the most robust diagnostic of late-type cool stars 
and brown dwarfs.

The sharp turnaround in Y-K and Y-H occurring in the T dwarfs stands 
in contrast to T dwarf colors at longer and shorter wavelengths.
K-L' colors appear to redden out to at least L7, 
become flat with advancing spectral type from L7 to T1, and then redden again 
from T1 to at least T6 (Stephens et al., 2001; Leggett et al. 2002).
However, hampering the utility of this color dignostic is that
L-band data on stellar/sub-stellar photospheres is hard to come by 
due to high thermal backgrounds and resulting low sensitivity.
I-z colors have been studied through only the L spectral types, with 
Steele \& Howells (2000) claiming that I-z colors redden through the M and
early L range, flatten from L1-L5, and then redden again towards even later 
spectral types.  However, I-band photometry is not readily obtainable 
for many of the nearby L and T objects due to source faintness (M$_I$ = 15
mag at M9 but $>$18 mag by L7).   
We thus advocate the use of Y-K, Y-H, and Y-J colors for identifying
very cool objects.

The downside of the Y-K vs Y-H diagram is the effect of extinction, which 
drives stars nearly directly along the dwarf locus.  In other color-color planes
such as I-Y or Y-J vs Y-K and J-K, however, there is more leverage of 
extinction vs the stellar/substellar color squence.

\section{Conclusions}

We present the first exploration of the colors of normal dwarf stars 
(and some brown dwarfs) spanning spectral types sdO through T5V 
using the Y-band at 1.035 $\mu$m.  
The behavior of the earth's atmosphere is relatively clean in this window 
of near unit transparency and few emission lines, 
especially compared to the other near-infrared J, H, and K windows.
Atmospheric extinction at Y-band is $<$0.05 mag per airmass.
The relationship between interstellar extinction in the Y-band
and that at V-band is $A_Y$ = (0.38$\pm$0.03) $\times A_V$.  

Our photometric results, including errors, appear in Table 2.  
Although for most objects only one measurement was taken,
photometric variability should not be an issue since our stars are mostly 
accepted photometric standards in either the optical or the near-infrared.
The exception to this are the late type dwarfs (late M, L, and T spectral types)
for which small-amplitude near-infrared variability has been reported
in some objects (e.g. Bailer-Jones \& Mundt, 2001).

The large range in color exhibited by low-mass stars and brown dwarfs
makes colors involving the Y-band particularly diagnostic
of these extremely cool objects.
The I-Y, Y-J, Y-H, and Y-K colors all grow redder with increasing
spectral type through at least L6V.  At later spectral types YJHK colors all
saturate and turn blueward to somewhat differing degrees, due to the strong 
peak in the SEDs of extremely cool objects in the 1.1-1.3 $\mu$m region. This
bluing is caused by H$_2$O, CH$_4$, and collision-induced H$_2$ opacity 
supressing first the K-band and then the H-band flux as temperatures grow cooler
(e.g. Marley et al. 2002).   Infrared spectroscopy of Gl229b and Gl570d, 
the coolest known T dwarfs,
appear to confirm these statements for most of the T spectral type range.
Colors involving the Y, J, H, and M bands can uniquely identify
extremely cool brown dwarfs and so-called ``planetary mass objects'' apart from 
other astronomical targets, and should form the basis for deep surveys
of field and cluster samples in the future.

\begin{acknowledgements}
We acknowledge with appreciation Adam Burgasser for obtaining P60/IRC (``X")
Y-band data for us of several T-dwarfs.
LAH thanks John Carpenter for allowing her to talk him into observing with
the Keck/NIRC (``Z") Y-band filter during data acquisition for our H/K study of 
Orion Nebula Cluster faint stars and brown dwarfs.  This publication  
makes use of data products from the Two Micron All Sky Survey (2MASS)
and has benefited from information provided by the SIMBAD database.  
\end{acknowledgements}

\clearpage
\begin{deluxetable}{rrr}
\tabletypesize{\normalsize}
\tablewidth{7truein}
\tablecaption{\bf Y-Band Filter Transmission \label{tab:filter}}
\tablehead{
\colhead{Wavelength [$\AA$]}   &
\colhead{Transmission (P60/IRC)}   &
\colhead{Transmission (Keck/NIRC)}   \\
}
\startdata
    9260. &0.0000 & 0.0000   \\
    9270. &0.0016 & 0.0000  \\
    9280. &0.0032 & 0.0000   \\
    9290. &0.0044 & 0.0000   \\
    9300. &0.0050 & 0.0000   \\
    9310. &0.0060 & 0.0000   \\
    9320. &0.0070 & 0.0010   \\
    9330. &0.0080 & 0.0014   \\
    9340. &0.0100 & 0.0024   \\
    9350. &0.0130 & 0.0026   \\
    9360. &0.0160 & 0.0037   \\
    9370. &0.0180 & 0.0051   \\
    9380. &0.0200 & 0.0059   \\
    9390. &0.0250 & 0.0078   \\
    9400. &0.0300 & 0.0097   \\
    9410. &0.0350 & 0.0117   \\
    9420. &0.0400 & 0.0140   \\
    9430. &0.0500 & 0.0170   \\
    9440. &0.0600 & 0.0211    \\
    9450. &0.0700 & 0.0259   \\
    9460. &0.0800 & 0.0300    \\
    9470. &0.1100 & 0.0365    \\
    9480. &0.1300 & 0.0462    \\
    9490. &0.1633 & 0.0542    \\
    9500. &0.2000 & 0.0618    \\
    9510. &0.2315 & 0.0758    \\
    9520. &0.2530 & 0.0847    \\
    9530. &0.2730 & 0.1082    \\
    9540. &0.3000 & 0.1264    \\
    9550. &0.3401 & 0.1505    \\
    9560. &0.3903 & 0.1716    \\
    9570. &0.4454 & 0.1944    \\
    9580. &0.5000 & 0.2192    \\
    9590. &0.5497 & 0.2539    \\
    9600. &0.5933 & 0.2948    \\
    9610. &0.6302 & 0.3248    \\
    9620. &0.6600 & 0.3938    \\
    9630. &0.6826 & 0.4533    \\
    9640. &0.6991 & 0.4982    \\
    9650. &0.7111 & 0.5455    \\
    9660. &0.7200 & 0.5871    \\
    9670. &0.7271 & 0.6177    \\
    9680. &0.7328 & 0.6435    \\
    9690. &0.7371 & 0.6621    \\
    9700. &0.7400 & 0.6748    \\
    9710. &0.7417 & 0.6864   \\
    9720. &0.7427 & 0.6976    \\
    9730. &0.7436 & 0.6999    \\
    9740. &0.7450 & 0.7057    \\
    9750. &0.7472 & 0.7090    \\
    9760. &0.7500 & 0.7109    \\
    9770. &0.7528 & 0.7124    \\
    9780. &0.7550 & 0.7136    \\
    9790. &0.7563 & 0.7151    \\
    9800. &0.7566 & 0.7152    \\
    9810. &0.7561 & 0.7165    \\
    9820. &0.7550 & 0.7165    \\
    9830. &0.7533 & 0.7165    \\
    9840. &0.7511 & 0.7165    \\
    9850. &0.7484 & 0.7165    \\
    9860. &0.7450 & 0.7134    \\
    9870. &0.7411 & 0.7125    \\
    9880. &0.7369 & 0.7110    \\
    9890. &0.7331 & 0.7083    \\
    9900. &0.7300 & 0.7058    \\
    9910. &0.7280 & 0.7019    \\
    9920. &0.7267 & 0.6993    \\
    9930. &0.7259 & 0.6964    \\
    9940. &0.7250 & 0.6935    \\
    9950. &0.7238 & 0.6891    \\
    9960. &0.7225 & 0.6841    \\
    9970. &0.7212 & 0.6816    \\
    9980. &0.7200 & 0.6770    \\
    9990. &0.7191 & 0.6739    \\
   10000. &0.7183 & 0.6708    \\
   10010. &0.7170 & 0.6655    \\
   10020. &0.7150 & 0.6628    \\
   10030. &0.7119 & 0.6614    \\
   10040. &0.7081 & 0.6574    \\
   10050. &0.7039 & 0.6547    \\
   10060. &0.7000 & 0.6534    \\
   10070. &0.6966 & 0.6534    \\
   10080. &0.6939 & 0.6493    \\
   10090. &0.6917 & 0.6480    \\
   10100. &0.6900 & 0.6467    \\
   10110. &0.6888 & 0.6426    \\
   10120. &0.6878 & 0.6426    \\
   10130. &0.6866 & 0.6426    \\
   10140. &0.6850 & 0.6427    \\
   10150. &0.6827 & 0.6427    \\
   10160. &0.6801 & 0.6427    \\
   10170. &0.6774 & 0.6373    \\
   10180. &0.6750 & 0.6359    \\
   10190. &0.6732 & 0.6319    \\
   10200. &0.6719 & 0.6292    \\
   10210. &0.6709 & 0.6279    \\
   10220. &0.6700 & 0.6265    \\
   10230. &0.6690 & 0.6239    \\
   10240. &0.6678 & 0.6225    \\
   10250. &0.6665 & 0.6225    \\
   10260. &0.6650 & 0.6225    \\
   10270. &0.6634 & 0.6212    \\
   10280. &0.6619 & 0.6198    \\
   10290. &0.6607 & 0.6172    \\
   10300. &0.6600 & 0.6172    \\
   10310. &0.6599 & 0.6145    \\
   10320. &0.6604 & 0.6131    \\
   10330. &0.6612 & 0.6105    \\
   10340. &0.6620 & 0.6105    \\
   10350. &0.6628 & 0.6091    \\
   10360. &0.6635 & 0.6091    \\
   10370. &0.6642 & 0.6078    \\
   10380. &0.6650 & 0.6078    \\
   10390. &0.6659 & 0.6092    \\
   10400. &0.6671 & 0.6092    \\
   10410. &0.6684 & 0.6078    \\
   10420. &0.6700 & 0.6092    \\
   10430. &0.6718 & 0.6092    \\
   10440. &0.6735 & 0.6092    \\
   10450. &0.6747 & 0.6092    \\
   10460. &0.6750 & 0.6105    \\
   10470. &0.6743 & 0.6119    \\
   10480. &0.6728 & 0.6119    \\
   10490. &0.6712 & 0.6146    \\
   10500. &0.6700 & 0.6146    \\
   10510. &0.6696 & 0.6146    \\
   10520. &0.6697 & 0.6133    \\
   10530. &0.6699 & 0.6133    \\
   10540. &0.6700 & 0.6146    \\
   10550. &0.6695 & 0.6146    \\
   10560. &0.6685 & 0.6133    \\
   10570. &0.6670 & 0.6133    \\
   10580. &0.6650 & 0.6120    \\
   10590. &0.6627 & 0.6093    \\
   10600. &0.6601 & 0.6079    \\
   10610. &0.6575 & 0.6066    \\
   10620. &0.6550 & 0.6026    \\
   10630. &0.6527 & 0.6026    \\
   10640. &0.6504 & 0.5999    \\
   10650. &0.6479 & 0.5972    \\
   10660. &0.6450 & 0.5959    \\
   10670. &0.6415 & 0.5945    \\
   10680. &0.6375 & 0.5945    \\
   10690. &0.6336 & 0.5892    \\
   10700. &0.6300 & 0.5865    \\
   10710. &0.6270 & 0.5865    \\
   10720. &0.6244 & 0.5865    \\
   10730. &0.6222 & 0.5838    \\
   10740. &0.6200 & 0.5798   \\
   10750. &0.6178 & 0.5757    \\
   10760. &0.6154 & 0.5744    \\
   10770. &0.6129 & 0.5690    \\
   10780. &0.6100 & 0.5636    \\
   10790. &0.6068 & 0.5623    \\
   10800. &0.6032 & 0.5582    \\
   10810. &0.5993 & 0.5569    \\
   10820. &0.5950 & 0.5542    \\
   10830. &0.5905 & 0.5529    \\
   10840. &0.5861 & 0.5435    \\
   10850. &0.5825 & 0.5408    \\
   10860. &0.5800 & 0.5408    \\
   10870. &0.5790 & 0.5367    \\
   10880. &0.5785 & 0.5341    \\
   10890. &0.5775 & 0.5260    \\
   10900. &0.5750 & 0.5233    \\
   10910. &0.5701 & 0.5193    \\
   10920. &0.5636 & 0.5166    \\
   10930. &0.5565 & 0.5139    \\
   10940. &0.5500 & 0.5072    \\
   10950. &0.5448 & 0.5004    \\
   10960. &0.5408 & 0.4991    \\
   10970. &0.5377 & 0.4924    \\
   10980. &0.5350 & 0.4910    \\
   10990. &0.5326 & 0.4870    \\
   11000. &0.5306 & 0.4843    \\
   11010. &0.5296 & 0.4803    \\
   11020. &0.5300 & 0.4776    \\
   11030. &0.5318 & 0.4736    \\
   11040. &0.5335 & 0.4736    \\
   11050. &0.5335 & 0.4695    \\
   11060. &0.5300 & 0.4668    \\
   11070. &0.5213 & 0.4669    \\
   11080. &0.5059 & 0.4669    \\
   11090. &0.4826 & 0.4642    \\
   11100. &0.4500 & 0.4628    \\
   11110. &0.4075 & 0.4588    \\
   11120. &0.3577 & 0.4480    \\
   11130. &0.3041 & 0.4426    \\
   11140. &0.2500 & 0.4292    \\
   11150. &0.1986 & 0.3942    \\
   11160. &0.1519 & 0.3713    \\
   11170. &0.1118 & 0.3349    \\
   11180. &0.0800 & 0.3147    \\
   11190. &0.0578 & 0.3038    \\
   11200. &0.0435 & 0.2929    \\
   11210. &0.0350 & 0.2819    \\
   11220. &0.0300 & 0.2710    \\
   11230. &0.0266 & 0.2601    \\
   11240. &0.0242 & 0.2492    \\
   11250. &0.0221 & 0.2383    \\
   11260. &0.0200 & 0.2274    \\
   11270. &0.0176 & 0.2164    \\
   11280. &0.0149 & 0.2055    \\
   11290. &0.0123 &  0.1946   \\
   11300. &0.0100 & 0.1837    \\
   11310. &0.0082&  0.1728   \\
   11320. &0.0069&  0.1618   \\
   11330. &0.0059&  0.1509   \\
   11340. &0.0050&  0.1400   \\
   11350. &0.0042&  0.1291   \\
   11360. &0.0034&  0.1182   \\
   11370. &0.0027&  0.1072   \\
   11380. &0.0020&  0.0963   \\
   11390. &0.0014&  0.0854  \\
   11400. &0.0008 & 0.0745    \\
   11410. &0.0004 & 0.0636    \\
   11420. &0.0000 & 0.0527    \\
   11430. &0.0000 & 0.0417    \\
   11440. &0.0000 & 0.0308    \\
   11450. &0.0000 & 0.0199    \\
   11460. &0.0000 & 0.0090    \\
   11470. &0.0000 & 0.0000    \\
\enddata
\end{deluxetable}

\clearpage

\begin{deluxetable}{ccrrrrrrrrrrrrl}
\tabletypesize{\scriptsize}
\tablewidth{10truein}
\rotate
\tablecaption{\bf Photometry of Optical/Infrared Standard Stars and Additional Red Objects\tablenotemark{1}   \label{tab:data}}
\tablehead{
\colhead{Source Name}   &
\colhead{SpT}   &
\colhead{B-V}   &
\colhead{V-I}   &
\colhead{I-K}   &
\colhead{Y}   &
\colhead{Y err.}  &
\colhead{Y-H}   &
\colhead{Y-H err.}  &
\colhead{Y-K}   &
\colhead{Y-K err.}  &
\colhead{J$_{2MASS}$}   &
\colhead{H$_{2MASS}$}   &
\colhead{K$_{2MASS}$}   &
\colhead{Comment}       \\

\colhead{}   &
\colhead{}   &
\colhead{[mag]}   &
\colhead{[mag]}   &
\colhead{[mag]}   &
\colhead{[mag]}   &
\colhead{[mag]}  &
\colhead{[mag]}  &
\colhead{[mag]}  &
\colhead{[mag]}  &
\colhead{[mag]}   &
\colhead{[mag]}  &
\colhead{[mag]}   &
\colhead{[mag]}  &
\colhead{}       \\
}
\startdata
PG 2213-006  &  sdO  & -0.217 & -0.203 & -0.378 & 14.559 & 0.030 & -0.132 & 0.048 & -0.167 & 0.108 &  14.70 & 14.78 & 14.60 &Landolt \\
HD 218902  &  A0V &-0.01&  -  &  -  & 10.244 & 0.036 & -0.011 & 0.045 & -0.020 & 0.046 & 10.23 & 10.22 & 10.20 & zero point\\
HD 222644  &  A0/A1V &0.1: &  -  &  -  & 10.571 & 0.032 &  0.010 & 0.040 &-0.004 & 0.040 &  10.56 & 10.54 & 10.54 & zero point\\
SA 112-805  &  A0/A1  & 0.152 & 0.138 & 0.205 & 11.836 & 0.030 & 0.051  & 0.038 & 0.065 & 0.041 &  -&-&-& Landolt / zero point\\
HD 191398  &  A0V &0.09 &  -  &  -  & 8.739 & 0.026 & 0.084 & 0.033 & 0.094 & 0.033 & 8.67 &8.60&8.61& zero point\\
HD 228700  &  A0V &0.11 &  -  &  -  & 8.905 & 0.026 & 0.119 & 0.033 & 0.126 & 0.034 & 9.392&9.33&9.31 &zero point\\
HD 161481  &  A0     &0.24 &  -  &  -  & 8.594 & 0.029 & 0.157 & 0.036 & 0.156 & 0.037 &  -&-&-&zero point\\
HD 161168  &  A0     &0.3: &  -  &  -  & 8.856 & 0.038 & 0.109 & 0.047 & 0.163 & 0.048 &  8.81&8.77&8.70&zero point\\
SA 109-71  &  A0 & 0.323 & 0.41 & 0.496 & 10.823 & 0.035 & 0.170 & 0.044 & 0.206 & 0.045&  -&-&-&Landolt / zero point\\
HD 167163  &  A0V &0.35 &  -  &  -  & 8.705 & 0.029 & 0.179 & 0.037 & 0.222 & 0.037 &  -&-&-&zero point\\
SA 115-420  &  F5  & 0.468 & 0.58 & 0.657 & 10.362 & 0.027 & 0.392 & 0.035 & 0.399 & 0.036 & -&-&-& Landolt \\
GSPC P576-F = SJ 9183 &  G     &  -  &  -  &  -  & 12.296 & 0.030 &  -                    &  -    & 0.419 &  -    & -&-&-& Persson \\
GSPC P550-C = SJ 9143\tablenotemark{3}  &  G  &  -  &  -  &  -  & 12.610 & 0.05  & 0.493 & 0.05  & 0.542 & 0.05  & -&-&-& Persson \\
SA 113-492  &  G0: & 0.553 & 0.684 & 0.785 & 11.279 & 0.028 & 0.545 & 0.036 & 0.546 & 0.038 & -&-&-& Landolt \\
GSPC P290-D = SJ 9188  &  G  &  -  &  -  &  -  & 11.875 & 0.026 & 0.524 & 0.034 & 0.581 & 0.036 & -&-&-& Persson \\
GSPC P330-E = SJ 9166  &  G  &  -  &  -  &  -  & 12.041 & 0.018 & 0.565 & 0.022 & 0.582 & 0.023 &  11.79&11.46&11.43&Persson \\
 FS 35   & K0                  & 0.8:&  -  &  -  & 12.359 & 0.030 & 0.52 \tablenotemark{5}    & 0.030 & 0.582 & 0.030 & 12.20\tablenotemark{4}&11.83\tablenotemark{4}& 11.75\tablenotemark{4}& Hawarden\\
GSPC S813-D = SJ 9182  &  G  &  -  &  -  &  -  & 11.697 & 0.009 & 0.552 & 0.010  & 0.586 & 0.010 &  11.47&11.12&11.13&Persson \\
GSPC P545-C = SJ 9134\tablenotemark{3}  &  G  &  -  &  -  &  -  & 12.155 & 0.05 & 0.531 & 0.05 & 0.598 & 0.05 & -&-&-& Persson \\
GSPC P182-E = SJ 9177  &  G  &  -  &  -  &  -  & 12.328 & 0.016 & 0.558 & 0.020 & 0.603 & 0.021 &  12.09&11.76&11.71&Persson \\
SA 113-493  &  G5:  & 0.786 & 0.824 & 0.97 & 10.666  & 0.028 & 0.608 & 0.036 & 0.664 & 0.037 &  -&-&-&Landolt \\
SA 109-949  &  G8: & 0.806 & 1.02 & 1.130 & 11.384 & 0.029 & 0.595 & 0.037 & 0.668 & 0.038 &  -&-&-&Landolt \\
GSPC S867-V = SJ 9155\tablenotemark{3}  &  G5 &0.68 &  -  &  -  & 12.363 & 0.05 & 0.719 & 0.05 & 0.729 & 0.05 & 11.98\tablenotemark{4}&11.64\tablenotemark{4}&11.59\tablenotemark{4}& Persson / Hawarden\\
2MASSI J2254188+312349&T5V     &  -  &  -  &  -  & 16.037 & 0.051 & 1.00 \tablenotemark{5}    & 0.051 & 0.823 & 0.092 & 15.28&15.04&14.83& late-type \\
Gl 229b\tablenotemark{2}  & T6.5V &  -  &  -  &$>$5.6  & 15.3~~~ & - & 1.0~~~ & - & 0.9~~~ & - &  -&-&-&late-type\\
SA 113-495  &  G:  & 0.947 & 1.01 & 1.306 & 11.055 & 0.028 & 0.849 & 0.036 & 0.906 & 0.037 &  -&-&-&Landolt \\
SA 110-477  &  -  & 1.345 & 1.707 & 1.685 & 11.572 & 0.036 & 0.823 & 0.045 & 0.947 & 0.046 & -&-&-& Landolt \\
SA 109-956  &  -  & 1.283 & 1.525 & 1.796 & 12.516  & 0.029 & 1.038 & 0.038 & 1.160 & 0.041 &  -&-&-&Landolt \\
Gl 754.1B  &  M2.5V & 1.469 & 2.87 & 1.475 & 8.547 & 0.037 & 0.964 & 0.045 & 1.193 & 0.045 & 9.54:&9.634&9.01:& late-type\\
SA 109-954  &  -  & 1.296 & 1.496 & 1.929 & 10.254 & 0.028 & 1.064 & 0.036 & 1.204 & 0.037 &  -&-&-&Landolt \\
LHS 3409 = GJ 4076  &  M3.5V-sdM4.5 & 1.85 & 2.66 & 2.223 & 11.486  & 0.027 & 0.964 & 0.034 & 1.211 & 0.035 & 10.98\tablenotemark{4} &10.49\tablenotemark{4}&10.26\tablenotemark{4}&late-type\\
GJ 1220 = LHS 3297  &  M4V & 1.73 & 2.96 & 2.489 & 10.128 & 0.032 & 1.083 & 0.040 & 1.339 & 0.040 &  -&-&-&late-type\\
GJ 1215 = LHS 3277  &  M5.5V & 1.96 & 3.32 & 2.875 & 10.377 & 0.032 & 1.105 & 0.040 & 1.404 & 0.040 &  -&-&-&  late-type\\
SA 110-364  &  -  & 1.133 & 1.281 & 1.996 & 11.775  & 0.035 & 1.120  & 0.043 & 1.409 & 0.044 & -&-&-& Landolt\\
LHS 523 = GJ 4281  &  M6.5V & 2.03 & 4.26 & 3.133 & 11.354 & 0.030  & 1.134 & 0.039 & 1.449 & 0.040 & 10.79\tablenotemark{4}&10.22\tablenotemark{4}&9.91\tablenotemark{4}&late-type\\
BRI 2202-1119 = LP 759-25 &  M6.5V &  -  &  -  & 2.936 & 12.226 & 0.032 & 1.158 & 0.040 & 1.464 & 0.040 & 11.68&11.06&10.73& late-type\\
TVLM 868-53850  &  M6:  &  -  &  -  &  -  & 12.195 & 0.031 & 1.197 & 0.039 & 1.504 & 0.040 & 11.53&10.94&10.65& Persson \\
VB 8  &  M7V & 2.2 & 4.56 & 3.405 & 10.455 & 0.033 & 1.232 & 0.042 & 1.591 & 0.042 & -&-&-& late-type\\
LHS 2632 = LP 321-222\tablenotemark{3} & M7.5V &  -  &  -  &  -  &13.001& 0.05 & 1.41\tablenotemark{5}  &  0.05 &  1.77\tablenotemark{5}~~  & 0.05 & 12.26&11.59&11.23& late-type\\
LHS 3003 = GJ 3877\tablenotemark{3}& M7V & 1.34&4.52 &3.196&10.735&0.05  & 1.41\tablenotemark{5}  &0.05 &  1.815  & 0.05 & 9.96 &9.32&8.92& late-type \\
2MASS J1524248+292535\tablenotemark{3} &  -& -  &  -  &  -  & 12.069 &0.05 & - &  -  & 1.847   & 0.05&  -  & -&-& late-type\\
SDSSp 175032.96+175903.9&T3.5  &  -  &  -  &  -  & 17.427 & 0.132 & 1.49 \tablenotemark{5}    & 0.132 & 1.96\tablenotemark{5}~~    & 0.132 &16.32&15.94&15.47& late-type; z*=19.63\\
SDSSp 015141.69+124429.6& T1   &  -  &  -  &  -  & 17.173 & 0.111 & 1.63 \tablenotemark{5}    & 0.111 & 1.965 & 0.138 & 16.25&15.54&15.18& late-type; z*=19.51\\
TVLM 868-110639  &  M9V &  -  &  -  & 4.422 & 13.470 & 0.033 & 1.584 & 0.042 & 2.064 & 0.043 & -&-&-& Persson \\
LHS 2924 = GJ 3849\tablenotemark{3}& M9V &1.62:&4.37 &4.446& 12.940 &0.05 & 1.77\tablenotemark{5}& 0.05  &  2.176  &0.05  & 11.92\tablenotemark{4}&11.17\tablenotemark{4}&10.75\tablenotemark{4}& late-type\\
2MASSW J1439284+192915\tablenotemark{3} & L1V &  -  &4.92 &4.53 &  13.778  &0.05 & 1.73\tablenotemark{5} &0.05 &  2.187  & 0.05& 12.76&12.05&11.58& late-type \\
2MASSW J0030438+313932  &  L2V &  -  &  -  &  -  & 16.448 & 0.033 & 1.864 & 0.047 & 2.418 & 0.078 & 15.49&14.58&13.99& late-type\\
DENIS-P J0205.4-1159&  L7V     &  -  &  -  &5.31 & 15.425 & 0.037 & 1.84 \tablenotemark{5}    & 0.037 & 2.437 & 0.042 & 14.58&13.59&12.98& late-type\\
2MASSW J2208136+292121& L2V    &  -  &  -  &  -  & 16.894 & 0.068 & 2.06 \tablenotemark{5}    & 0.068 & 2.826 & 0.076 & 15.82&14.83&14.09& late-type\\
2MASSW J1728114+394859& L7V    &  -  &  -  &  -  & 16.821 & 0.064 & 2.04 \tablenotemark{5}    & 0.064 & 2.92\tablenotemark{5}~~  & 0.064 & 15.96&14.78&13.90& late-type\\
2MASSI J1029216+162652\tablenotemark{3} &L2.5& -  &  -  &  -  & 15.565  &0.05 & 2.22\tablenotemark{5} &0.05 &  2.952  & 0.05&  14.31&13.35&12.61& late-type\\
2MASSW J1632291+190441  &  L8V &  -  &  -  &  -  & 16.984 & 0.043 & 2.260 & 0.059 & 2.993 & 0.081 & 15.86&14.59&13.98& late-type\\
2MASSI J0103320+193536& L6V    &  -  &  -  &  -  & 17.312 & 0.123 & 2.43 \tablenotemark{5}    & 0.123 & 3.169 & 0.127 & 16.26&14.88&14.15& late-type\\
L 134-5  &  -  &  -  &  -  &  -  & 12.976 & 0.033 & 3.296 & 0.041 & 4.012 & 0.041 &  -&-&-&Persson / extincted \\
``Oph2" = 2MASS J162705-242010  &  -  &  -  &  -  &  -  & 17.530 & 0.094 & 3.677 & 0.105 & 4.612 & 0.109 &  12.70&10.50&9.36&extincted \\
``Oph5" = 2MASS J162705-242036  &  -  &  -  &  -  &  -  & 14.813 & 0.038 & 3.724 & 0.047 & 4.668 & 0.047 &  -&14.47&12.97&extincted \\
L 547  &  -  &  -  &  -  &  -  & 13.750 & 0.030 & 3.833 & 0.037 & 4.741 & 0.037 &  10.83&10.39&10.25&Persson / extincted \\
``Oph9" = 2MASS J162703-241832  &  -  &  -  &  -  &  -  & 18.408 & 0.118 & 3.647 & 0.137 & 4.907 & 0.136 &  -&-&-&extincted \\
``Oph8" = 2MASS J162707-242009  &  -  &  -  &  -  &  -  & 14.582 & 0.038 & 4.117 & 0.046 & 5.186 & 0.046 &  -&-&-&extincted \\
``Oph6" = 2MASS J162706-241923  &  -  &  -  &  -  &  -  & 19.248 & 0.309 & 4.202 & 0.320 & 5.363 & 0.321 &  -&14.86&13.83&extincted \\
``Oph11" = 2MASS J162658-241837  &  -  &  -  &  -  &  -  & 18.758 & 0.167 & 4.120  & 0.179 & 5.363 & 0.179 &  -&14.51&13.34&extincted \\
``Oph3" = 2MASS J162702-242040  &  -  &  -  &  -  &  -  & 17.519 & 0.111 & 4.518 & 0.117  & 5.883 & 0.117 &  -&-&-&extincted \\ 
\enddata

\tablenotetext{1}{
Table is sorted by Y-K color.
Spectral types and luminosity classes are given where available and taken, along
with B-V and V-I photometry from the literature. I-K colors are calculated by 
combining literature I with the present K measurements.
Y, Y-H, and Y-K photometry is newly measured here.  
Errors include terms for zero point, atmospheric extinction,
and photon statistics added in quadrature. J$_{2MASS}$, H$_{2MASS}$, and 
K$_{2MASS}$ values come from 2MASS' Second Incremental Release and have
errors typically 0.03 - 0.04 mag.  Comment indicates the type of object,
either a photometric standard star from Landolt (1992), Persson (1998),
or Hawarden (2001), a later type usually spectroscopic standard star, or 
an extincted/reddened object.
}
\tablenotetext{2}{
Gl 229b photometry is from Nakajima et al. 1995 and Matthews et al., 1996. 
}
\tablenotetext{3}{
YHK data is from Keck/NIRC instead of P60/IRC. 
}
\tablenotetext{4}{
Star's JHK photometry is from Geballe et al 2002, Hawarden et al. 2001, 
Leggett et al. 1998, or Leggett 1992 (largely in MKO/UKIRT system) 
instead of from 2MASS.
}
\tablenotetext{5}{
Star's quoted Y-H or Y-K color is from measured Y magnitude 
and 2MASS or other H,K/K$_s$ photometry.
}

\end{deluxetable}

\clearpage  

\clearpage


\begin{references}


Bahng, J. 1969, MNRAS 143, 73

Bailer-Jones, C.A.L. \& Mundt, R. 2001, AA 367, 218

Barsony, M., Kenyon, S.J., Lada, E.A., \& Teuben, P.J., 1997 ApJS 112, 109

Bessell, M.S. \& Brett, J.M.  1988, PASP 100, 1134

Burgasser, A.J., Kirkpatrick, J.D., Brown, M.E., Reid, I.N., Burrows, A.,
Liebert, J., Matthews, K., Gizis, J.E., Dahn, C.C., Monet, D.G., Cutri, R.M.,
\& Skrutskie, M.F. 2002, ApJ 564, 421

Burrows, A., Hubbard, W.B., Lunine, J.I., \& Liebert, J.
2001, Rev. Mod. Phys, in press. 

Burrows, A., Marley, M., Hubbard, W.B., Lunine, J.I., Guillot, T., Saumon, D.,
Freedman, R., Sudarsky, D., \& Sharp, C. 1997, ApJ 491, 856

Cohen, J.G., Persson, S.E., Elias, J.H., \& Frogel, J.A. 1981, ApJ 249, 481 

Fukugita, M., Ichikawa, T., Gunn, J.E., Doi, M., Shimasaku, K., Schneider, D.P.
1996 AJ 111, 1748 

Geballe, T.R., Knapp, G.R., Leggett, S.K., Fan. X., et al 2002, ApJ 564, 466

Hawarden, T.G., Leggett, S.K., Letawsky, M.B., Ballantyne, D.R., \& Casali,
M.M. 2001, MNRAS, 325, 563

Hawley, S.L., Gizis, J.E., \& Reid, I.N. 1996, AJ 112, 2799

Hillenbrand, L.A. \& Carpenter, J.M. 2000, ApJ 540, 236

Johnson, H.L. 1966, ARAA 4, 193

Kirkpatrick, J.D., Henry, T.J., \& McCarthy, D.W. 1991, ApJS 77, 417

Kirkpatrick, J.D., Henry, T.J., \& Simons, D.A. 1995, AJ 109, 797

Kirkpatrick, J.D., Reid I.N., Liebert J., Cutri R.M., Nelson B., Beichman C.A., Dahn C.C., Monet D.G., Gizis J.E., Skrutskie M.F.  1999, ApJ 519, 802

Kirkpatrick, J.D., Reid, I.N., Liebert, J., Gizis, J.E., Burgasser, A.J.,
Monet, D.G., Dahn, C.C., Nelson, B., \& Williams, R.J, 2000, AJ 120, 447

Landolt, A.U. 1992, AJ 104, 340

Lasker, B.M., Sturch, C.R., Lopez, C., Mallama, A.D., et al., 1988, ApJS 68, 1

Leggett, S.K. 1992, ApJS 82, 351

Leggett, S.K., Allard, F., \& Hauschildt, P.H. 1998, ApJ 509, 836

Leggett, S.K., Allard, F., Dahn, C., Hauschildt, P. H., Kerr, T. H., 
\& Rayner, J.  2000a ApJ 535, 965

Leggett, S.K., Geballe, T.R., Fan, X., Schneider, D.P., et al. 2000b ApJL 536, 35

Leggett, S.K., Golimowski, D.A., Fan, X., Geballe, T.R., et al. 
2002, ApJ 564, 452 

Manduca, A. \& Bell, R.A. 1979, PASP 91, 848

Marley, M.S., Saumon, D., Guillot, T., Freedman, R.S., Hubbard, W.B., 
Burrows, A., \& Lunine, J.I. 1996, Science 272, 1919

Marley, M.S., Seager, S., Saumon, D., Lodders, K., Ackerman, A.S., \& Freedman, R. 2002 ApJ, 568, 335

Matthews, K., \& Soifer, B. T. 1994, Infrared Astronomy with Arrays: the Next Generation, ed. I. McLean (Dordrecht: Kluwer), 239 

Matthews, K., Nakajima, T., Kulkarni, S.R., \& Oppenheimer, B.R. 1996, AJ 112, 1678

Murphy, D.C., Persson, S.E., Pahre, M.A., Sivaramakrishnan, A.,M. \& Djorgovski, S.G. 1995, PASP 107, 1234 

Nakajima, T., Oppenheimer, B.R., Kulkarni, S.R., Golimowski, D.A., 
Matthews, K., \& Durrance, S.T. 1995, Nature 378, 463

Reid, I.N., Hawley, S.L., \& Gizis, J.E. 1995, AJ 110, 1838

Persson, E., Murphy, D.C., Krzeminski, W., Roth, M., \& Rieke, M.J. 1998, AJ 116, 2475

Schlegel, D.J., Finkbeiner, D.P., \& Davis, M. 1998, ApJ 500, 525

Selby, J.E.A. \& McClatchey, R.A., ``Atmospheric Transmittance From 
0.25 to 28.5 $\mu$m: Computer Code LOWTRAN 3,'' Air Force
Cambridge Research Laboratories, 7 May 1975

Steele, I.A., \& Howells, L. 2000, MNRAS 313, L43

Stephens, D.C., Marley, M.S., Noll, K.S., \& Chanover, N. 2001, ApJL 556, 97

Tinney, C.G. 1993, ApJ 414, 279

Tokunaga, A.T., Simons, D.A., \& Vacca, W.D. 2002, PASP 114, 180

\end{references}
\end{document}